\documentclass[letterpaper]{article}

\usepackage{aaai}
\usepackage{times}
\usepackage{helvet}
\usepackage{courier}
\usepackage{threeparttable}
\usepackage{booktabs, dcolumn}

\usepackage{graphicx}
\usepackage{array,booktabs,ragged2e}
\usepackage{hyperref}
\usepackage{xcolor}
\hypersetup{
colorlinks=true,
		pdfborder={0 0 0},
	citecolor=gray,
	linkcolor=red,
	urlcolor=blue}
\usepackage[tight,footnotesize]{subfigure}

\usepackage{color}
\usepackage{amsmath}
\usepackage{amsfonts}
\usepackage{epsfig}
\usepackage{url}
\usepackage{multirow}
\usepackage[ruled,vlined,linesnumbered]{algorithm2e}
\usepackage{setspace}
\usepackage{wrapfig}

\newcommand{\gar}{{\textsc{GlobalAR}}}
\newcommand{\lar}{{\textsc{LocalAR}}}
\newcommand{\sdar}{{\textsc{SDAR}}}

\newcommand{\itunes}{{\textsc{SWM}}}
\newcommand{\flip}{{\textsc{Flipkart}}}

\newcommand{\bit}{\begin{itemize}}
\newcommand{\eit}{\end{itemize}}
\newcommand{\ben}{\begin{enumerate}}
\newcommand{\een}{\end{enumerate}}
\newcommand{\beq}{\begin{equation}}
\newcommand{\eeq}{\end{equation}}

\newcommand{\hide}[1]{}

\title{Temporal Opinion Spam Detection by Multivariate Indicative Signals
	}

\author{
	\makebox[80pt]{Junting Ye \quad\quad\quad Santhosh Kumar  \quad\quad\quad\quad Leman Akoglu}\\
	{Stony Brook University}\\
	{Department of Computer Science}\\
	{\{juyye,smanavasilak,leman\}@cs.stonybrook.edu}\\
}

\begin{document}

\maketitle

\vspace{-3in}

\begin{abstract}
	Online consumer reviews reflect the testimonials of real people, unlike advertisements.
	As such, they have critical impact on potential consumers, and indirectly on businesses. According to a Harvard study \cite{Luca2011}, +1 rise in star-rating increases revenue by 5--9\%.
	Problematically, such financial incentives have created a market for spammers to fabricate reviews, to unjustly promote or demote businesses, activities known as opinion spam \cite{JindalL08}. 
	A vast majority of existing work on this problem have formulations based on \textit{static} review data, with respective techniques operating in an \textit{offline} fashion.
	Spam campaigns, however, are intended to make most impact during their course. Abnormal events triggered by spammers' activities could be masked in the load of future events, which static analysis would fail to identify. 
	In this work, we approach the opinion spam problem with a \textit{temporal} formulation.
	Specifically, we monitor a list of carefully selected indicative signals of opinion spam over time and design \textit{efficient} techniques to both detect \textit{and} characterize abnormal events in \textit{real-time}. 
	Experiments on datasets from two different review sites show that our approach is fast, effective, and practical to be deployed in real-world systems.
\end{abstract}

\section{Introduction}
\label{sec:intro}
Online product reviews play important role for e-commerce. New customers tend to prefer products with higher ratings as previous buyers  have ``testified'' that the products are good choices. On the contrary, new customers may have less interest for products with lower ratings since more dissatisfaction from experienced users have been reported. Driven by such commercial benefits, spam or fake reviews have become a prevalent problem, for which effective detection algorithms are greatly needed \cite{JindalL08}.

In the past years, several existing works employed supervised techniques by extracting features based on review text, ratings, product meta-data such as category and price, review feedback such as number of helpful votes, etc. \cite{JindalL08,Ott11,FengACL12A,conf/kdd/MukherjeeKLWHCG13}. These methods have a key challenge as ground truth is extremely hard to obtain---human annotators can hardly tell the difference between the genuine and fake reviews \cite{Ott11}. 

Unsupervised approaches have also been explored for the opinion spam problem.
In a nutshell, these can be categorized as approaches that leverage 
linguistic
\cite{Ott11,FengACL12A},
relational
\cite{WangXLY11,conf/icwsm/AkogluCF13,conf/icdm/LiCLWS14,ye2015discovering}, 
and most often, behavioral clues
\cite{conf/cikm/JindalLL10,conf/icwsm/FengXGC12,conf/www/MukherjeeLG12,lim2010detecting,conf/kdd/MukherjeeKLWHCG13,xie2012review}.
In their investigation, \cite{conf/icwsm/MukherjeeV0G13} found that Yelp's fake review filter might be relying more on behavioral rather than text-based clues.
Most recently, \cite{conf/kdd/RayanaA15} proposed a unifying approach that harnesses all of these information sources, which outperforms any individual one alone, especially when provided with a few labeled instances.

As one can see, previous work has focused on aspects such as availability of labels or the lack thereof, or the type of information to consume (linguistic, relational, behavioral). 
Orthogonal to these aspects, in this work we bring emphasis on the aspect of \textit{time}. 
Surprisingly, we realize that the vast majority of past work has formulated the opinion spam problem on {\em static} review data, and developed techniques that operate in an {\em offline} fashion.
A few works, such as those by \cite{conf/icwsm/FeiMLHCG13} and \cite{conf/icwsm/LiCM0S15}, designed and used temporal features along with static ones. Nevertheless, their approaches are effectively offline where the entire review data is assumed to be available at analysis time.
Recently, \cite{xie2012review} proposed an approach that tracks three temporal features---average rating, number of reviews, and ratio of singleton reviewers---over time, and pinpoints time windows in which all three change simultaneously. Their approach is specifically crafted for catching \textit{singletons}, i.e., spam reviewers with only a {single} review, that earlier methods mainly failed to effectively identify.

In this work, we consider opinion spam as a \textit{temporal} phenomenon.
Our intuition is that the spammers' activities trigger abrupt changes and unusual events in the underlying generating process {\em at the time} of their operations. 
Such changes may later be reverted, or the impact of such events may be lost in the abundance of future events.
As such, we argue that a temporal, \textit{real-time} approach would not only be more effective but also more realistic, as review data is temporal in nature.
In the following we provide an overview of our proposed methodology, and the list of contributions.

\subsection{Overview}
To promote or demote products, spammers need to dominate the sentiments among a given product's reviews, which requires considerable amount of reviews. Such intensive behaviors result in abnormal bursts of various suspicious activities over the timeline of the product. In this work, our goal is twofold: we aim to \textit{both} detect \textit{and} characterize those abnormal bursts of activities. 

In a nutshell, we first identify a number of signals associated with suspicious reviewing behavior. For each product (or business), these signals are tracked over time in an online fashion.
Through efficient time series modeling, we detect anomalous time points among all the signals.
Finally, we leverage the specific anomalous signals for characterization, i.e., inspection and sensemaking.
We outline the main steps of our approach as follows:

\vspace{-0.025in}
\begin{enumerate}
	\setlength{\itemsep}{-0.75\itemsep}
	\item {\em Temporal signal extraction:} 
	We define a list of measures, called \textit{indicative signals}, that provide potential evidence toward opinion spam.
	These measures are computed for each product over time as new reviews arrive.
	As such, each product is represented by multiple time series.
	
	\item {\em Anomaly detection in lead (target) signal:} %(\S \ref{ssec:lead}) 
	We designate one of the signals as \textit{the lead}. 
	Anomalous changes in the lead signal suggest potential spamming activity. 
	
	\item {\em Anomaly detection in supporting signals:} %(\S \ref{ssec:others}) 
	Anomalies in the lead signal
	provide the necessary conditions for opinion spam; but not necessarily sufficient. 
	For endorsement, we perform anomaly detection on the remaining signals in an efficient way, by focusing on time periods around the anomalous points (``alarms'') detected in the lead.
	
	\item {\em Characterization:}  %(\S \ref{ssec:pscore}) 
	Our method spots anomalous time points across all signals for each product. We use a function of the number and magnitude of the anomalies to rank the products. At any given time, products are  investigated in their ranked order. The specific detected time points and  support signals guide the manual inspection.
	
\end{enumerate}
\vspace{-0.065in}

\subsection{Contributions}
\begin{figure}[!t]
	\centering
	\begin{tabular}{c}
		\hspace{-0.1in}	\includegraphics[width=3.15in,height=4.65in]{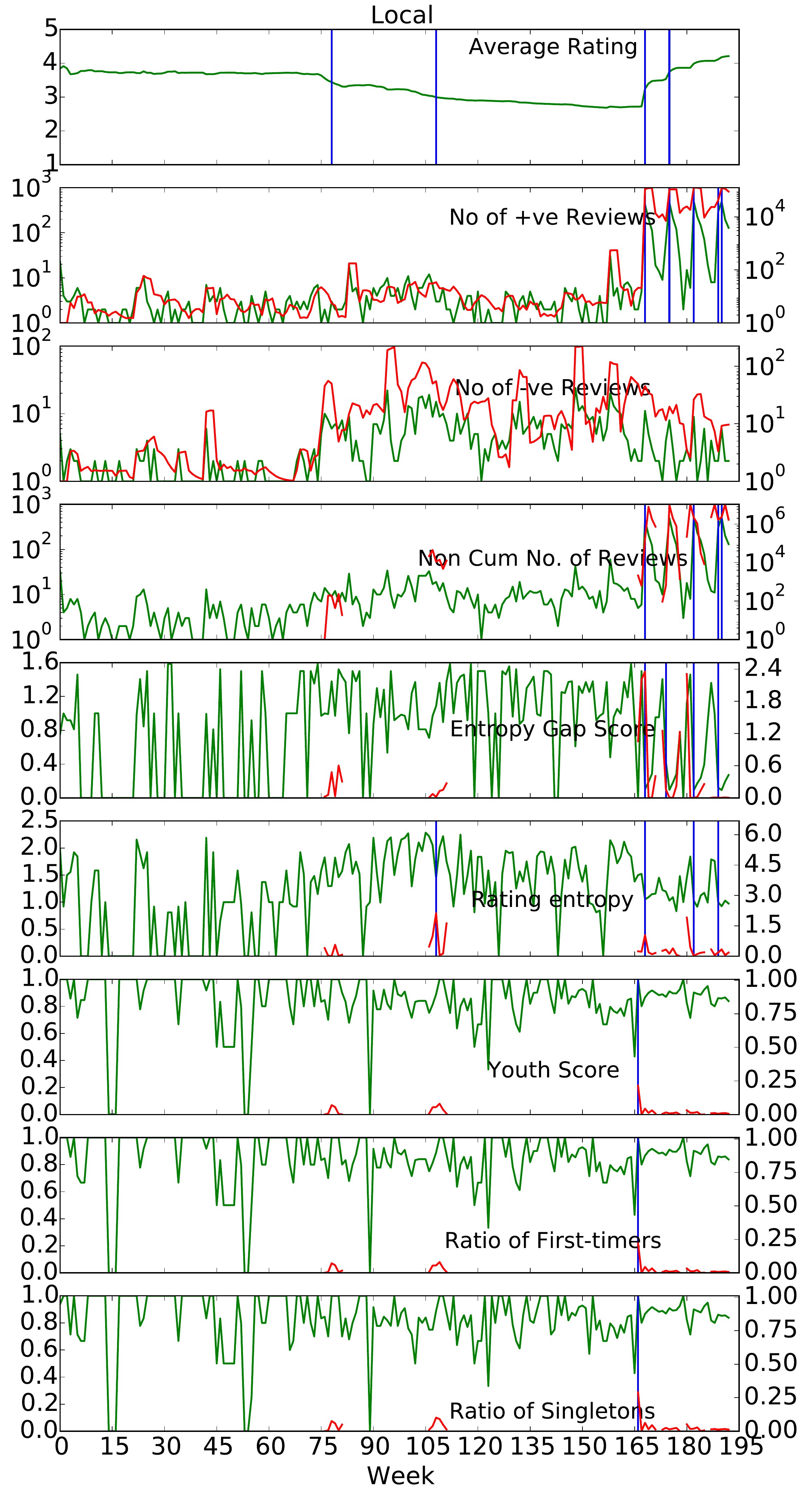} \\
	\end{tabular}
	\vspace{-0.175in}
	\caption{\em \small
		Time series for 9 indicative signals for a software product (app) (green curves: signal values, red curves: anomaly score, blue bars: anomalous time points). Our method detects 4 weeks of attacks that increase the average rating of a product with generally declining rating. Results discussed further in experiments.
	} 
	\label{fig:crown}
	\vspace{-0.2in}
\end{figure}

\vspace{-0.015in}
\begin{itemize}
	\setlength{\itemsep}{-0.5\itemsep}
	\item  \textbf{Problem formulation}: 
	We propose a new temporal approach to the opinion spam problem, that monitors a  carefully selected list of indicative signals.
	Anomalies across various signals suggest and corroborate evidence. 
	As shown in the experiments, our approach is 
	($i$) \textit{descriptive}; it facilitates inspection and sensemaking, and 
	($ii$) \textit{general}; it can identify different kinds of spam campaigns.
	
	\item \textbf{A new methodology}: We develop an \textit{online} and \textit{efficient} algorithm for opinion spam.
	We mainly monitor the lead signal and detect anomalies in \textit{real-time}. Only when an ``alarm'' is raised in the lead, a \textit{local} algorithm then performs anomaly detection in the support signals by considering a small window around the anomaly (hence local).
	
	\item \textbf{Practicality}: The proposed approach is usable in practice, as it provides capabilities for online monitoring, automatic detection, and interpretability for the analysts. 
	Specifically, detected time points narrow down where to look, and the indiciative signals suggest what to look at. We demonstrate the practicality of our approach through case studies in two review datasets, for apps in an online marketplace and for products in Flipkart (cf. Figure \ref{fig:crown}).
\end{itemize}

\noindent
\textbf{Reproducability}: Open source code of our method is available at:
{\small{\url{http://www3.cs.stonybrook.edu/~juyye/code/ICWSM16_Code.zip}}}.

\vspace{-0.1in}
\section{Indicative Signals of Opinion Spam}
\label{sec:signals}

We consider the review data of a product $p$ as a stream of review units ordered in time, $D_p = \{u_1,u_2,...,u_i,...\}$. Each unit consists of user ID, timestamp, review text or content, and rating; i.e, $u_i = (UID_i,t_i,c_i,r_i)$, where $t_i \leq t_j$ if $i<j$. Timestamps are divided into fixed-length time windows ($\Delta T$) where the values of each time series are computed based on data from these intervals.

We identify eight types of indicative signals, and extract their time series for each product $p$. 
A list of these signals are given in Table \ref{tab:signals} which we describe as follows.
\vspace{-0.075in}
\ben
\item \textbf{Average Rating}: 
If the average rating of a product $p$ changes significantly, it might have been tempered with by spammers.
This time series tracks the evolution of \textit{cumulative} average rating.
Let $\bar{U}_p^t = \{u_k | t_k \in [0, t*\Delta T)\}$ denote the set of $p$'s reviews until the end of time window $t$, where $|\bar{U}_p^t|=m$. Then,

\vspace{-0.25in}
\begin{equation}
	{R}_p^t = \frac{1}{m}\sum_{k=1}^m r_k
\end{equation}

\begin{threeparttable}[!t]
	\hspace{-0.15in}
	\vspace{-0.15in}
	\begin{center}
		\caption{\em Indicative signals of opinion spam. 	\label{tab:signals}}
		\begin{tabular}{l|r|r}
			\hline
			\hspace{-0.15in}		\textbf{Name}  & \textbf{Range}  & \textbf{Suspicious if}  \\ \hline
			\hspace{-0.15in}		Average Rating & $[1,5]$ & Change \\
			\hspace{-0.15in}		Number of ($+$/$-$) Reviews & $[0,\infty]$ & Increase \\
			\hspace{-0.15in}		Rating Entropy & $[0,\log_25]$ & Decrease \\
			\hspace{-0.15in}		Ratio of Singletons & $[0,1]$ & Increase \\
			\hspace{-0.15in}		Ratio of First-timers & $[0,1]$ & Increase \\
			\hspace{-0.15in}		Youth Score & $[0,1]$ & Increase \\
			\hspace{-0.15in}		Temporal Gap Entropy & $[0,maxe$\tnote{\textdagger}$\;\;]$ & Decrease \\
			\hline
		\end{tabular}
		\begin{tablenotes}
			\item[] {	\hspace{-0.2in}\textdagger \small{With windows size $\Delta T$ and logarithmic binning, 
					number of bins is $\lceil log_2\Delta T\rceil+1$ and
					$maxe=\log_2(\lceil log_2\Delta T\rceil+1)$.}}
		\end{tablenotes}
	\end{center}
	\vspace{0.1in}
\end{threeparttable}

\item \textbf{Number of Reviews}: A sudden surge in the number of reviews may indicate spamming activity. As such, we track the total number of reviews within each time interval. Let $U_p^t = \{u_k | t_k \in [(t-1)*\Delta T, t*\Delta T)\}$ denote $p$'s reviews within window $t$. Then,

\vspace{-0.15in}
\begin{equation}
	C_p^t = \big| U_p^t \big|
\end{equation}

\item \textbf{Number of Positive/Negative Reviews}: 
In addition to total number of reviews, we also track the positive and negative review counts, as fake reviews either demote or promote and have skewed ratings.
\begin{equation}
	+C_p^t = \big| \{u_k | u_k\in U_p^t, r_k\in \{4,5\}\} \big|
\end{equation}
\vspace{-0.15in}
\begin{equation}
	-C_p^t = \big| \{u_k | u_k\in U_p^t, r_k\in \{1,2\}\} \big|
\end{equation}

\item \textbf{Rating Entropy}: We also monitor entropy over time, as a measure of skewness in ratings. As such,
\vspace{-0.1in}
\begin{equation}
	{E_p^t} = -\sum\limits_{r=1}^{5} p_r^t\cdot\log p_r^t 
\end{equation}
where $p_r^t$ is the ratio of reviews with rating value equal to $r$ in window $t$.

\item \textbf{Ratio of Singletons}: Fake reviews could be posted by newly created accounts. Therefore, 
we track the ratio of one-time reviewers over time.

\vspace{-0.1in}
\begin{equation}
	{S_p^t} = \frac{|U_s^t|}{C_p^t}
\end{equation}
where $U_s^t$ is the set of singleton users who posted their first-ever \textit{and only} review (to $p$) during window $t$.

\item \textbf{Ratio of First-timers}: Some campaigns involve spammers targeting multiple products simultaneously. As such, we also track the ratio of first-time reviewers as

\vspace{-0.1in}
\begin{equation}
	{F_p^t} = \frac{|U_f^t|}{C_p^t}
\end{equation}
where $U_f^t$ is the set of $p$'s reviewers who posted their first but not necessarily only review during window $t$ (note that in this case the first review need not be to $p$).

\item \textbf{Youth Score}: 
Most fake reviews are posted from short-lived accounts. Therefore, besides singletons and first-timers, we further track account age.
Specifically, for each review $u_k$ of $p$ within window $t$, we compute the age of its reviewer $UID_k$ at the time they posted $u_k$ by
$$
A_k  = t_k-t_0^{UID_k}
$$
where $t_0^{UID_k}$ is the time at which reviewer $UID_k$ posted their first review.
The youth score is then the average of reviewer ages at the time they posted for $p$.

\vspace{-0.1in}
\begin{equation}
	Y_p^t  = \frac{1}{C_p^t} \;\sum\limits_{u_k \in U_p^t} 2\cdot(1-\frac{1}{1+\exp{(-A_k)}})
\end{equation}
\vspace{-0.1in}

The larger the $A_k$'s, the closer $Y_p^t$ gets to zero. As such, larger $Y_p^t$ is more suspicious.

\item \textbf{Temporal Gap Entropy}: 
A normal user is expected to post reviews at arbitrary intervals, while spammers may follow a predictable robot-like behavior (e.g., post every day). 
As such, we compute the time-gap between consecutive reviews $\{u_k, u_{k+1}\} \in U_p^t$ of $p$ within window $t$, create a histogram,
and compute the entropy. 
The histogram is constructed through logarithmic binning \cite{Newman2005PowerlawsPareto} (e.g., if $\Delta T$ is 30 days, consecutive bin sizes are 0, 1, 2, 4, 8, and 16 days). %, where $B$ denotes the number of bins.
That is,

\vspace{-0.125in}
\begin{equation}
	{G_p^t} = -\sum\limits_{b=1}^{\lceil log_2\Delta T\rceil+1} p_b^t\cdot \log p_b^t 
\end{equation}
where $p_b^t$ is the ratio of time-gaps in bin $b$ in window $t$.

\een

Notice that the temporal signals of interest are potential indicators of spamming behavior.
However, they need not exhibit abnormalities altogether at the same time windows.
In fact, different spam campaigns might trigger different signals to fire.
For example, one campaign may create a large number of new accounts but use them in arbitrary time intervals.
Another may involve a very skewed number of positive ratings but from sufficiently old accounts. 
Our approach is multi-faceted, as such it could help identify various combinations of collusions among spammers.

\vspace{-0.05in}
\section{Temporal Opinion Spam Detection}
\label{sec:change}

Our temporal approach to opinion spam detection consists of four main steps: \textit{(i)} extracting temporal signal values as defined in the previous section; \textit{(ii)} detecting changes in what is called the \textit{lead} signal; \textit{(iii)} checking whether we also observe temporal anomalies in the supporting signals; and \textit{(iv)} ranking targeted products based on the number and magnitude of anomalies found in their timeline.

In the following subsections, we describe the details of the last three steps, respectively.

\vspace{-0.05in}
\subsection{Anomalies in the Lead (Target) Signal}
\label{ssec:lead}

Out of all the indicative signals that we extract over time, we dedicate one of them as the {\em lead} signal.
The lead can be chosen as the measure that spammers particularly aim to manipulate, e.g., the average rating. It can also be a measure for which spamming activities could trigger a change, e.g., the number of (positive, negative, or total) reviews.

To elaborate, average rating serves as the overall impression of consumers for a given product.
It has the potential to shape the first impression of candidate consumers.
Therefore, many spammers are devoted to manipulate the average rating of products \cite{lim2010detecting}.
If there is a sudden change in average rating over time (either increase or decrease), it is possible that the product is a target of a (promotion or demotion) spam campaign.

On the other hand, it gets harder and harder to change the average rating for products with increasingly many reviews, compared to those with only a few. 
In their analysis, \cite{rahman2015catch} show that the minimum number of reviews needed to fraudulently increase the average rating of a product by 1/2 star is $n$/7, where $n$ is the number of genuine reviews of the product.
This suggests that it is costlier to change the average rating of popular (i.e., famous or notorious) products with large number of reviews, provided spammers are paid per review \cite{paid11}. 
Nevertheless, as we show in the experiments, there exist scenarios where we observe a burst of fake reviews in the timeline of a product, even though those do not have any impact on the average rating.
The incentives can be various; such as promoting a new feature of a product, flooding the most recent page view (as reviews are often ordered in time), etc.
In order to spot such spam campaigns, we use number of positive (or negative) reviews as the lead signal and watch for abnormal increases.

For anomaly detection on the lead time series, one can use any detection algorithm that provides real-time capability. One important issue is the semantics of the lead signal.
Average rating is cumulative, and our goal is to find \textit{change points} in both directions (either increase or decrease). For this lead, we use the cumulative sum (CUSUM) change detection algorithm \cite{Page3720621}.
On the other hand, we track the non-cumulative number of ($+$ or $-$) reviews per time window $\Delta T$, with a goal to spot \textit{anomalies} in the form of bursts (i.e., large increases). For such leads, we use the autoregressive (AR) model for modeling the lead series, and use the deviation from the forecasted value as the anomaly score $s$ at a new time point. 

For change/anomaly detection, choice of a threshold is critical to flag alerts. 
For a given (lead) signal and a detector, we maintain the distribution $\mathcal{D}(S|T,P)$ of the anomalousness scores $S$ ($i$) across time points $T$ and ($ii$) across products $P$. We then employ Cantelli's inequality\footnote{Unlike the t-test which assumes Gaussian data, Cantelli's inequality does not make any distributional assumptions on $\mathcal{D}$.} to identify a theoretical threshold $\delta = (\mathcal{D},\eta)$, where $\eta$ is the expected percentage of anomalies (see \cite{smets2011odd}).  For a given score $s_p^t$ for product $p$ at time $t$,
we flag an alert if $s_p^t > \delta$ and the anomaly is in the direction of what a suspicious activity would create (e.g., increase in youth score but decrease in entropy, see Table \ref{tab:signals}).

\vspace{-0.05in}
\subsection{Anomalies in the Supporting Signals}
\label{ssec:others}

\begin{table}[t!]
	\vspace{-0.1in}
	\begin{center}
		\caption{\em Notation used throughout text.}
		\vspace{0.01in}
		\label{tab:notations}
		\small{
			\begin{tabular}{p{0.4in}|p{2.5in}}
				\hline  \textbf{Notation} &  \textbf{Definition}\\ 
				\hline  $v_{i-k}^{i-1}$  &   a sequence of input values $(v_{i-k},...,v_{i-1})$\\
				\hline  $t_i$  &  $i_{th}$ time point\\
				\hline  $t_a$  &  most recent alarm time point\\
				\hline  $k$  &  order of AR (autoregressive) model\\
				\hline  $r$  &  discounting parameter of \sdar\\
				\hline  $\delta$  &  anomalousness score threshold\\
				\hline  $\boldsymbol{\omega}$  &  AR coefficients\\
				\hline  $\mu$  &  mean of time series values\\
				\hline  $\sigma$  &  variance of time series values\\
				\hline  $\boldsymbol{\theta}$  &  $\boldsymbol{\theta} = \{\boldsymbol{\omega},\mu,\sigma\}$\\
				\hline  $s_{i-4}^{i-1}$  &  a sequence of square errors ($s_{i-4},...,s_{i-1}$)\\
				\hline  $O_{i-4}^{i-1}$  &  a sequence of anomaly labels ($O_{i-4},...,O_{i-1}$)\\
				\hline  $L$  &  number of data points used to pick $k$\\
				\hline 
			\end{tabular} 
		}
	\end{center}
	\vspace{-0.25in}
\end{table}

Lead signal alone is not sufficient to indicate the occurrence of spamming activities. For example, when a sudden increase is detected in lead signal, e.g. average rating, the causes could be \textit{(i)} the product owner hired spammers to write fake reviews for promotion; \textit{(ii)} the product's quality has improved such that genuine reviewers wrote more positive reviews.  Therefore, we investigate further the supporting signals, to verify if ``alarms'' triggered by the lead signal are indeed the consequences of spamming activities. 

In order to detect anomalies in each supporting signal, we first propose \gar, which is an \textit{online} algorithm that quantifies the extent an input value $v_i$ is an anomaly in time series setting. Its major component is the Sequentially Discounting Auto Regression algorithm (\sdar) \cite{yamanishi2002unifying}. \sdar~ detects temporal anomalies by modeling a given time series with autoregressive model and discounting the importance of historical data exponentially. 

\textsc{GlobalAR}~ is effective but computationally expensive as it computes the anomaly score for each and every time point in the supporting signals. Therefore, we modify \gar~ and propose \lar~ to reduce time complexity. Given a supporting signal, \lar~ only focuses on and scores the time points around the ``alarms'' produced by the lead signal, and hence selectively ignores the other time points. This reduces the time complexity from linear to \textit{sublinear}, in terms of the total number of time points. In the following subsections, we elaborate on the technical details of \gar~ and \lar, respectively.

\vspace{-0.05in}
\subsubsection{\gar}
\gar~ detects temporal anomalies by using \sdar. Simply put, \sdar~ predicts value (denoted as $\hat{v}_i$) at time point $t_i$ and computes square error between prediction and observation, i.e., $(v_i - \hat{v}_i)^2$. If the square error is large, then $v_i$ is more likely to be an anomaly. It is easy to see that the square error can be large when $\hat{v}_i$ is very large or very small. However, in the setting of opinion spam detection, only either one of the cases indicates the occurrence of spamming activities. For example, a product might have been attacked by spammers if the rating entropy drops significantly (i.e. reviewers during the same time point give very similar ratings). In contrast, if rating entropy suddenly increases, it does not necessarily indicate the activities of spammers. In the remainder of text, we call $v_i$ as ``semantically suspicious'' (denoted as $SemSus(v_i) = True$) if (1) $v_i > v_{i-1}$ and large value indicates spamming activities (e.g., ratio of singletons), or (2) if $v_i < v_{i-1}$ and small value indicates suspicious behaviors (e.g., rating entropy). Otherwise $SemSus(v_i) = False$.

Next, we introduce the general idea of \sdar. It first models time series $V = \{v_k,v_{k+1},...,v_n\}$ with $k_{th}$ order autoregressive model as in Equation \ref{eqa:zEqua} and \ref{eqa:vEqua}. 

\vspace{-0.05in}
\begin{equation}\label{eqa:zEqua}
	z_i = \boldsymbol{\omega}z_{i-k}^{i-1} + \epsilon
\end{equation}

\vspace{-0.1in}
\begin{equation}\label{eqa:vEqua}
	v_i = z_i + \mu
\end{equation}
where $z_{i-k}^{i-1} = (z_{i-k},...,z_{i-1})$, $\boldsymbol{\omega} = (\omega_1,...,\omega_k)$, $\epsilon$ is a normal random variable with zero mean and $\sigma$ variance, i.e. $\epsilon \sim \mathcal{N}(0, \sigma)$, $\mu$ is the mean value of the time series. 

Then the probability density function of $v_i$ is defined as Equation \ref{equ:likelihood}.

\vspace{-0.25in}
\begin{equation}\label{equ:likelihood}
	p(v_i|v_{i-k}^{i-1},\boldsymbol{\theta}) = \frac{1}{(2\pi)^{1/2}\sigma} \exp(-\frac{(v_i-\hat{v}_i)^2}{2\sigma^2})
\end{equation}
where $\hat{v}_i = \boldsymbol{w}(v_{i-k}^{i-1}-\boldsymbol{\mu})+\mu$, $v_{i-k}^{i-1} = (v_{i-k},...,v_{i-1})$, $\boldsymbol{\mu}$ is a $k$-dimensional vector with all elements equal to $\mu$, and $\boldsymbol{\theta} = \{\boldsymbol{\omega},\mu,\sigma\}$. 

\sdar~ discounts the importance of historical data and estimates  $\boldsymbol{\theta}$ by maximizing the exponentially weighted log likelihood as shown in Equation \ref{eqa:objFunc}.

\vspace{-0.25in}
\begin{equation}\label{eqa:objFunc}
	L(\boldsymbol{\theta}|v_1^{i},r,k) = \sum\limits_{m=k+1}^{i} (1-r)^{i-m} \log p(v_m|v_{m-k}^{m-1},\boldsymbol{\theta})
\end{equation}
where $r\in [0,1)$ is the discounting factor.

In Equation \ref{eqa:objFunc}, the estimation of $\boldsymbol{\theta}$ is defined on the entire time series. Fortunately, it can also be estimated incrementally. Due to limited space, we skip the technical details of the incremental estimation (See \cite{yamanishi2002unifying} for details). We use $\boldsymbol{\theta_i} = \sdar(v_i,\boldsymbol{\theta_{i-1}}, r, k)$ to denote the update of parameters from $\boldsymbol{\theta_{i-1}}$ to $\boldsymbol{\theta_{i}}$. 

\begin{algorithm}[t!]
	% 	\setstretch{0.8}
	\DontPrintSemicolon
	\SetAlgorithmName{Algorithm}{procedure}{List of procedures}
	\caption{\gar} \label{alg:GAR}
	% 	\SetAlgoNlRelativeSize{-4}
	\small
	\textbf{Input:} $t_i$, $v_{i-k}^{i}$, $\boldsymbol{\theta_{i-1}}$, $k$, $r$, $t_a$,  $\delta$ \; %, $s_{i-4}^{i-1}$, $O_{i-4}^{i-1}$\;   
	\textbf{Output:} $\boldsymbol{\theta_i}$, $s_{t_a-2}^i$, $O_{t_a-2}^i$\; %{i-4}^i$  \;
	$\boldsymbol{\theta_i} = $\sdar$(v_i,\boldsymbol{\theta_{i-1}}, r, k)$ (where $\boldsymbol{\theta_i} = \{\boldsymbol{\omega_i},\mu_i,\sigma_i\}$)\;
	$\hat{v}_i = \boldsymbol{\omega_i}(v_{i-k}^{i-1} - \boldsymbol{\mu_i}) + \mu_i$ \;
	$s_i = (v_i -\hat{v}_i)^2$ \;
	%	Init $O_i = 0$ \;
	\If({\em // skip if no recent anomaly in lead signal}){$t_i - t_a \leq 2$}{
		\ForEach({// check time points around $t_a$}){$t_j \in [t_a-2, t_i]$}{
			\If{$s_j > \delta$ \& SemSus($v_j$) is $True$}{
				$O_j = 1$ \;
			}
		}
	}
\end{algorithm}

We show the steps of \gar~ in Algorithm \ref{alg:GAR}. We first estimate the parameters of \sdar, given a new input value $v_i$ (Line 3). Note that only $r$ and $k$ are hyper-parameters. $\boldsymbol{\theta_{i-1}}$ is the parameters estimated in the last iteration. Then we predict $\hat{v}_i$ using the estimated parameters $\boldsymbol{\theta_{i}}$ (Line 4). If $v_i$ is significantly different from  $\hat{v}_i$ (measured by square error $s_i$), then $v_i$ is deemed suspicious compared to past data (Line 5). In Lines 6 to 9, we first check if the current time point $t_i$ is close to the most recent ``alarm'' $t_a$ from the lead signal. If so, we  investigate the time points around $t_a$, particularly  $[t_a-2, t_i]$, instead of simply looking at $t_i$. The reason is that there might be lags between the lead and the supporting signals. For every time point $t_j\in [t_a-2, t_i]$, the anomaly label $O_j$ is assigned 1 if the corresponding square error $s_j$ is larger than threshold $\delta$ and $SemSus(v_j)$ returns $True$. 

\begin{algorithm}[t!]
	\DontPrintSemicolon
	\SetAlgorithmName{Algorithm}{procedure}{List of procedures}
	\caption{\lar} \label{alg:LAR}
	\small
	\textbf{Input:} $t_i$, $t_a$, $L, v_{i-L-5}^{i}$,  $\delta$ \;   
	\textbf{Output:} $s_{t_a-2}^i$, $O_{t_a-2}^i$  \;
	\If({{\em // exit if no recent anomaly in lead signal}}){$t_i - t_a > 2$ }{Exit}
	\ForEach({{\em // select $k$ that minimizes square error}}){$k' \in [1, 5]$}{
		Init $S_{k'} = 0$ \;
		\ForEach{$j \in [1, L]$}{
			$\boldsymbol{\theta} = AR(v_{i-j-k'}^{i-j-1}, k')$ (where $\boldsymbol{\theta} = \{\boldsymbol{\omega},\mu,\sigma\}$) \;
			$\hat{v}_{i-j} = \boldsymbol{\omega}(v_{i-j-k'}^{i-j-1} - \boldsymbol{\mu}) + \mu$ \;
			$s_{i-j} = (v_{i-j} -\hat{v}_{i-j})^2$ \;
			$S_{k'} = S_{k'} + s_{i-j}$ \;
		}
	}
	$k = k'_{min}$, where $\forall k'\in [1,5], S_{k'} \geq S_{k'_{min}}$ \;
	%	{Init} $O_{t_a-2}^i = 0$ \;
	\ForEach({{\em // check time points around $t_a$}}){$t_j \in [t_a-2, t_i]$}{
		$\boldsymbol{\theta_j} = AR(v_{j-k}^{j-1}, k)$\;
		$\hat{v}_j = \boldsymbol{\omega_j}(v_{j-k}^{j-1} - \boldsymbol{\mu_j}) + \mu_j$ \;
		$s_j = (v_j -\hat{v}_j)^2$ \;
		\If{$s_j > \delta$ \& SemSus($v_j$) is $True$}{
			$O_j = 1$ \;
		}
	}
\end{algorithm}
\setlength{\textfloatsep}{0.05in}

\subsubsection{Local AR}
In \gar, we investigate all the values in a time series. However, we expect and find that anomalies in the lead signal are much fewer compared to the number of time points. Recall that \gar~ exponentially discounts the importance of historical data. It implies that at time point $t_i$, the values that are much earlier than $t_i$ make little impact when estimating the parameters $\boldsymbol{\theta_i}$. Since we are mainly interested in time points ``alarmed'' by the lead signal, the question becomes if it would influence the performance much if we only focused on time points close to the ``alarms''. Motivated by this, we propose \lar. As we show in our experiments, both algorithms perform similarly but \lar~ is significantly faster than \gar.

\lar~ detects anomalies by using the autoregressive (AR) model. Different from \sdar, AR estimates parameters $\boldsymbol{\theta}$ by maximizing the data log likelihood $p(v_i|v_{i-k}^{i-1},\boldsymbol{\theta})$ (see Equation \ref{equ:likelihood}). The data likelihood is  dependent only on $v_{i-k}^{i-1}$, such that we can ignore values that are far away from $v_i$. Besides the improvement of efficiency, we pick a proper order, i.e. $k$, for AR by fitting regular (non-anomalous) values in the time series. We expect that estimating $k$ would be more effective than simply fixing $k$, as generally, the behaviors of time series from different products are not the same. Similar to \gar, we allow lags between the lead and supporting signals in \lar.

Algorithm \ref{alg:LAR}~ shows the detailed steps of \lar. At time $t_i$, we check if there is a close-by ``alarm'' $t_a$ in the lead signal. If not, then we exit the algorithm (Lines 3-4). This step accelerates the algorithm significantly in two aspects: ($i$) it skips anomaly score computation for points away from the lead ``alarms'', as a result of which ($ii$) feature extraction for a large body of time points in supporting signals can also be skipped. Next, we select a proper (integer) $k$ that minimizes the total square error over a window of $L$ values\footnote{Window size $L$  is essentially the training data size used to estimate $k$. Choice of $L$ poses a trade-off between the estimation quality  and running time. In experiments we find $L=8$ effective.} before $v_i$ (Lines 5-12). Specifically, we pick a candidate $k'$ and initialize the square error sum $S_{k'}$ to 0. We compute the square error $s_{i-j}$ between the inputs $v_{i-j}$ and predictions $\hat{v}_{i-j}$ of an AR model of order $k'$, for all $L$ values before $v_i$. Sum of square errors is denoted by $S_{k'}$. We then choose the $k'$ with the minimum $S_{k'}$ as the order of our AR model at time $t_i$. 
As temporal dependence drops by distance in time, we focus on small $k'\in [1,5]$.
Through Lines 13-18, we carefully examine the time points around $t_a$. Different from \gar, we compute the square error for values at and before $v_i$ at this step, using the estimated $k$. When the square error is larger than the anomaly threshold and $SemSus(v_j)$ returns $True$, anomaly label $O_j$ is assigned 1.

\vspace{-0.05in}
\subsection{Scoring and Ranking Products}
\label{ssec:pscore}
For a product $p_i$ at time point $t_j$, we can use CUSUM, \gar~ or \lar~ to detect anomalies in both its lead and supporting signals. Given these anomalies, how can we quantify the suspiciousness of $p_i$ at $t_j$? How can we rank $p_i$ among all other products at that time? 

We answer these questions by formulating a suspiciousness score for the products. Intuitively, it is more probable that a product is a target of spamming activities \textit{(i)}  if there are a large number of temporal anomalies among its indicative signals, and
\textit{(ii)} if the magnitudes of the  anomalies are large. 
Based on these insights, we design four measures to quantify product suspiciousness.

First is the ratio of anomalies among product $p_i$'s nine indicative signals at $t_j$.
That is, $f_1(p_i,t_j) = \sum_{l=1}^9 O^{(l)}_{j, p_i}/9$,
where $O^{(l)}_{j, p_i} \in \{0,1\}$ is the anomaly label of signal $l$ of product $p_i$ at $t_j$.
The second and third measures are respectively the average and maximum magnitude of the anomalies, and can be written as
$f_2(p_i,t_j) = \sum_{l=1}^9  s^{(l)}_{j, p_i} / \sum_{l=1}^9 O^{(l)}_{j, p_i}$ and 
$f_3(p_i,t_j) = \max_{l=1\ldots 9} s^{(l)}_{j, p_i}$, 
where $s^{(l)}_{j, p_i}$ is the anomaly score of signal $l$ of product $p_i$ at $t_j$.
Finally, $f_4(p_i,t_j) =  \sum_{l=1}^9 w_l \cdot s^{(l)}_{j, p_i}$ is the weighted sum of the anomaly scores, where
$w_l = 1 / \sum_{t=1}^j O^{(l)}_{t, p_i}$.
Simply put, the weight $w_l$ is inverse proportional to the number of anomalies in signal $l$. 
The intuition is that the signals that contain a large number of anomalies may be noisy and provide us with many potential false positives.

We then use the empirical CDF to normalize the feature values, as shown in Equation \ref{equ:cdf}.
\begin{equation}\label{equ:cdf}
	F_g(p_i,t_j) = P(f_g \leq f_g(p_i,t_j)), \;\; g=1, \ldots, 4
\end{equation}
where $F_g(p_i,t_j)$ is the ratio of $f_g$'s that are smaller than or equal to $f_g(p_i,t_j)$ across all products and all time points before $t_j$.
The larger the $F_g(p_i,t_j)$, the larger the anomalousness.
Finally, we compute $p_i$'s suspiciousness score $A(p_i,t_j)$ at time $t_j$ as an average of the $F_g(p_i,t_j)$'s.

\section{Experiments}
\label{sec:experiments}

\subsection{Datasets}
In this study we use review datasets from two different sites, namely \itunes~and \flip, as we describe next.

\textit{SoftWare Marketplace} (\itunes) consists of reviews for all software products (apps) from the entertainment category (e.g., games, movies, sports, etc.) in an anonymous online marketplace which allows customers
to purchase software applications.  It contains 15,094 apps with over 1.1 million reviews by 966 thousand
users, and spans 198 weeks between July 2008 and April 2012.
The data for this marketplace was originally collected and used by \cite{conf/icwsm/AkogluCF13}.

\flip~contains reviews
from flipkart.com, an e-commerce site which provides
a platform for sellers to market products to customers. It
contains 545 thousand products with roughly 3.3 million reviews by 1.1 million users,
and spans 180 weeks between August 2011 to January
2015.

\vspace{-0.05in}
\subsection{Results}
Both the \itunes~and the \flip~datasets do not contain
ground truth labels for anomalous (i.e., opinion-spammed) products, let alone the specific time periods in which each product was spammed.
As such, we manually inspect the top-ranked products from both datasets, and provide evidence through various case studies. 

\vspace{-0.015in}
\subsubsection{\itunes~Case I: Game app}
We start with the further analysis and description of the product shown in Figure \ref{fig:crown}.
This app, which started off with an average rating of 4, gradually declined below 3-star rating between weeks 75 to 165.
A series of spam campaigns are then executed in weeks
168, 175, 182, and 189. Notice that these campaigns are organized every 7 weeks, which is a strong indication of manipulation.

When we use `Average Rating' as the lead signal, we spot the first two campaigns, whereas when `No of + Reviews' is used as the lead, all 4 weeks are detected (note the blue bars indicating the time points with anomaly score above the threshold). Nearly all the supporting signals also change simultaneously, suggesting low rating entropy and increased number of singleton reviewers.

\begin{figure}[h]
	\vspace{-0.1in}
	\centering
	\begin{tabular}{cc}
		\hspace{-0.15in}	\includegraphics[width=1.7in,height=0.8in]{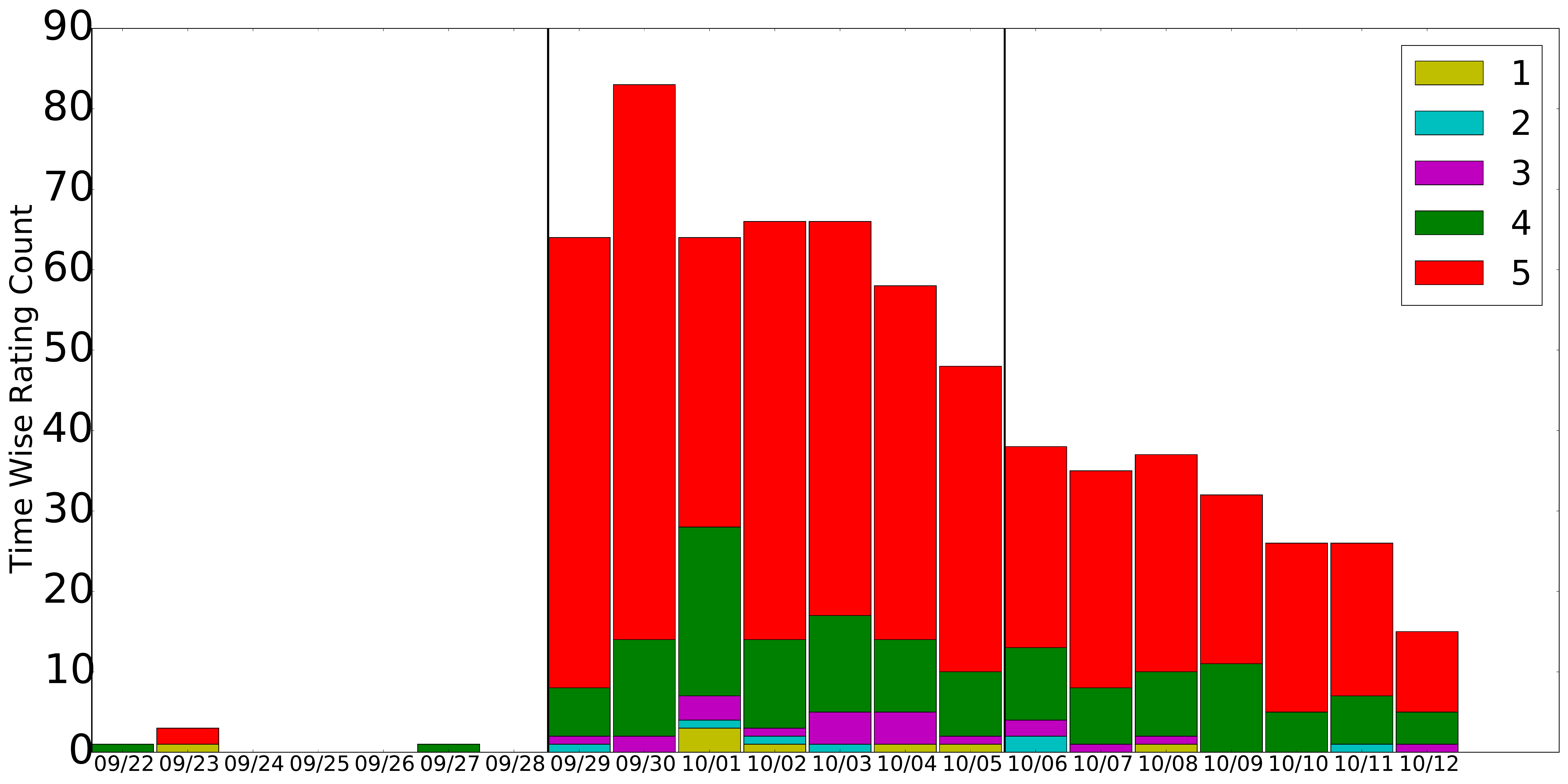} &
		\hspace{-0.2in}	\includegraphics[width=1.7in,height=0.8in]{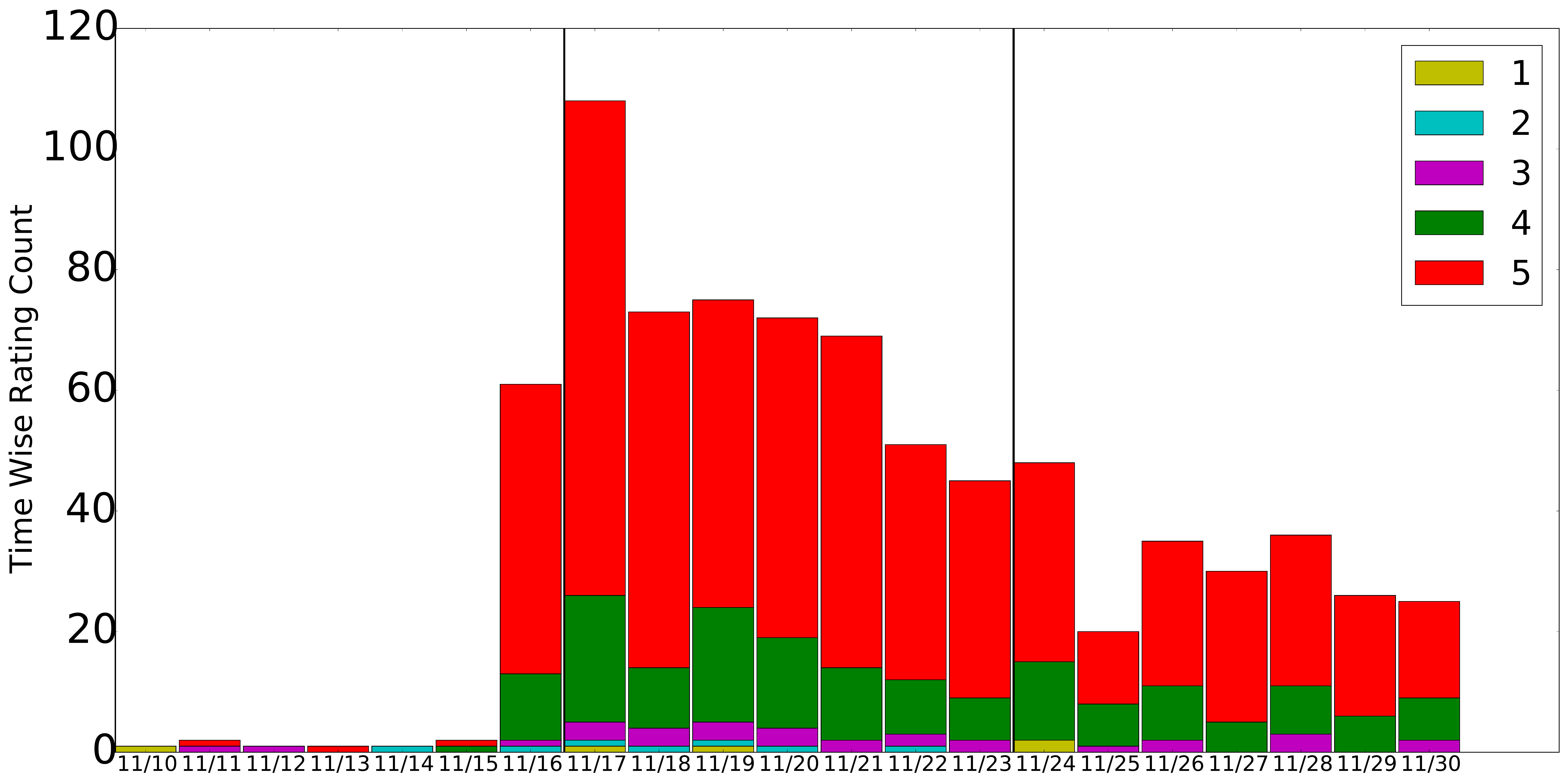} \\
		\hspace{-0.15in}	\includegraphics[width=1.7in,height=0.8in]{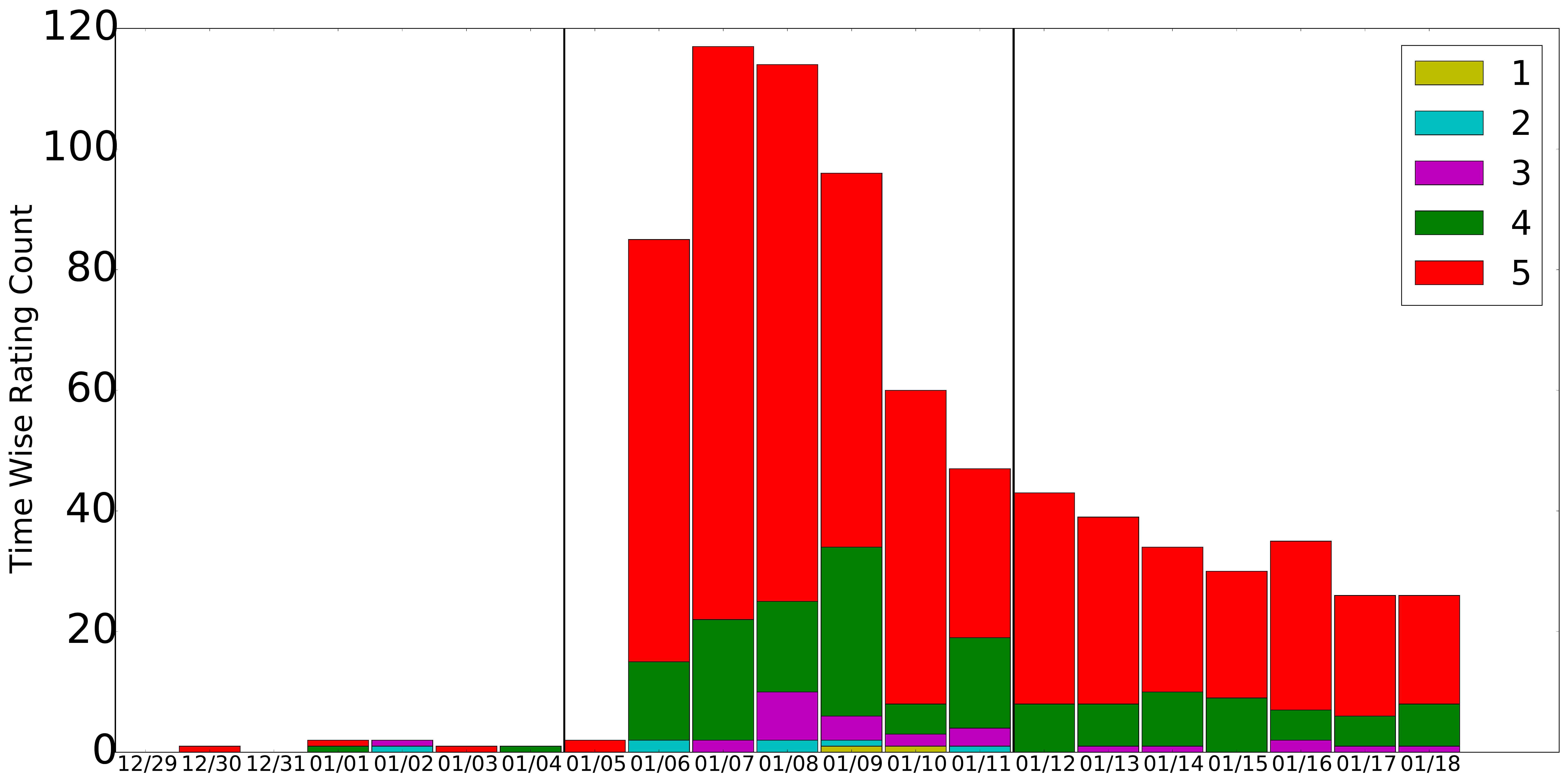} &
		\hspace{-0.2in}	\includegraphics[width=1.7in,height=0.8in]{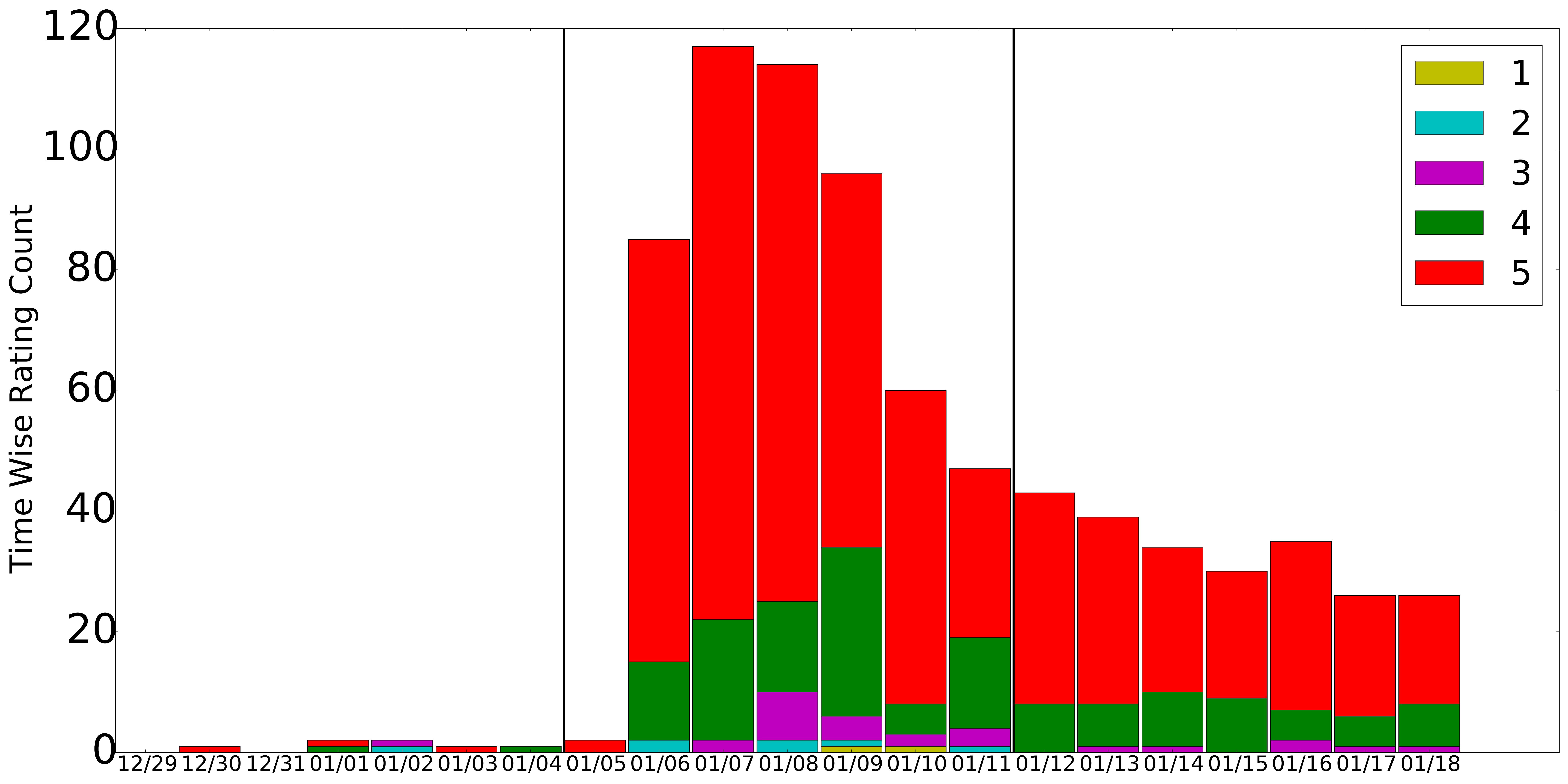} \\
	\end{tabular}
	\vspace{-0.175in}
	\caption{\em \small
		Stacked bar charts showing daily review counts for the 4 detected campaigns in Figure \ref{fig:crown} (week before, during, and after campaign separated by green vertical bars).
		Stacks represent counts for different ratings 1-5. 
		Notice that spam campaigns involve mostly 5-star reviews (hence the bumpy increase in average rating week by week after each campaign).
	} 
	\label{fig:swmcase1}
	%	\vspace{-0.1in}
\end{figure}

\begin{figure}[h]
	\centering
	\begin{tabular}{cc}
		\hspace{0.0in}	\includegraphics[width=1.3in,height=1.3in]{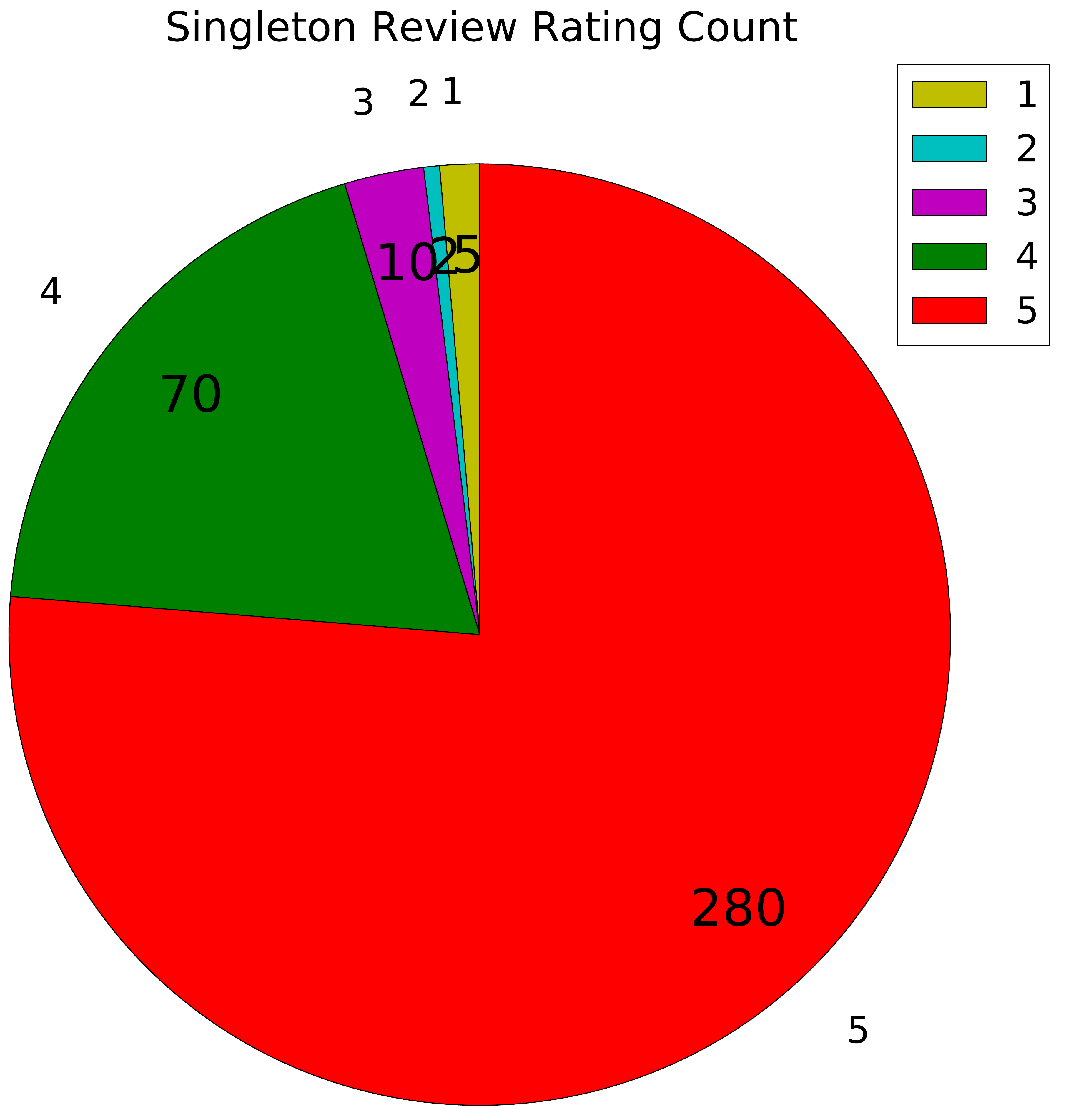} &
		\hspace{0.1in}	\includegraphics[width=1.3in,height=1.3in]{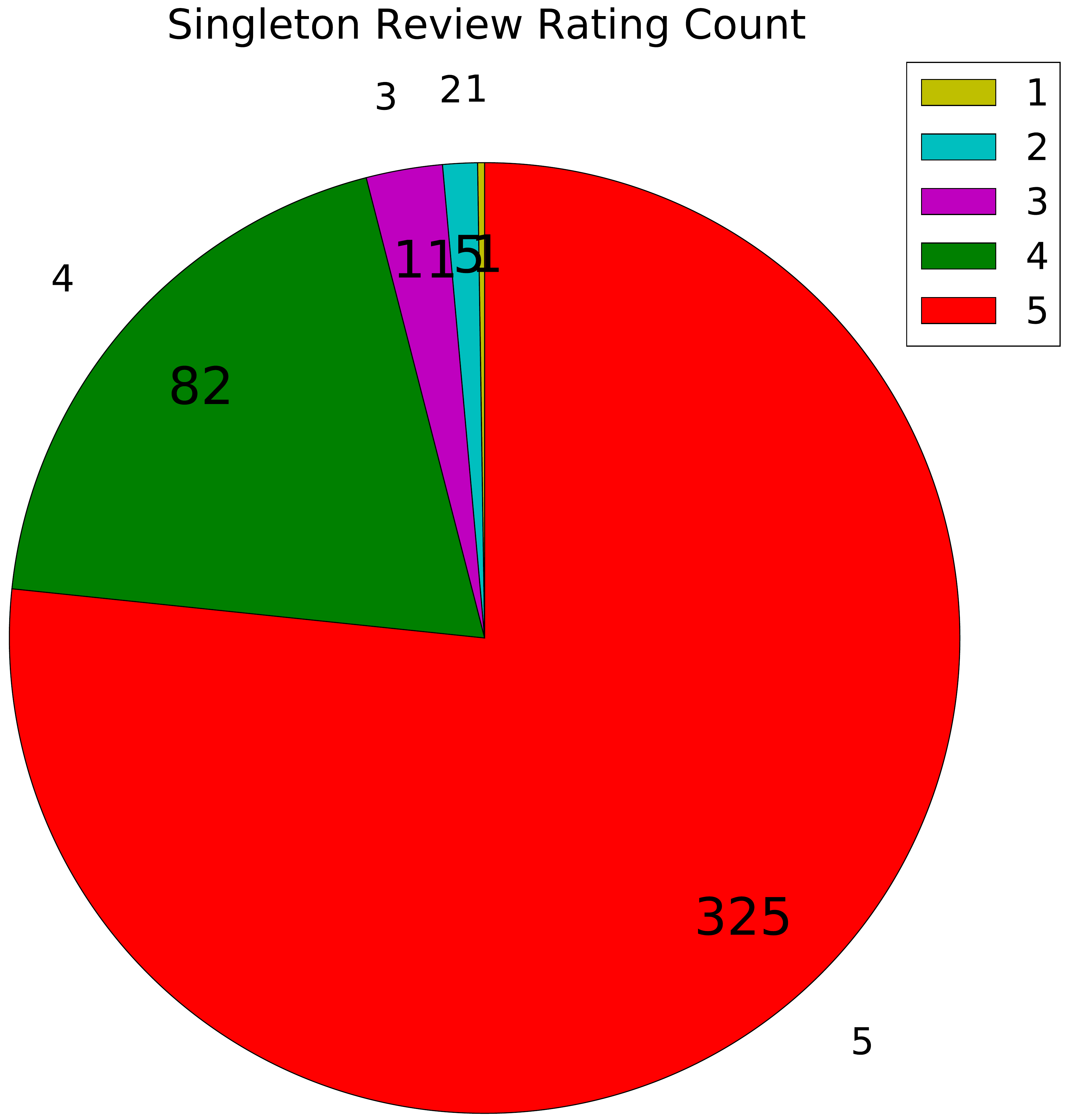} \\
		\hspace{0.0in}	\includegraphics[width=1.3in,height=1.3in]{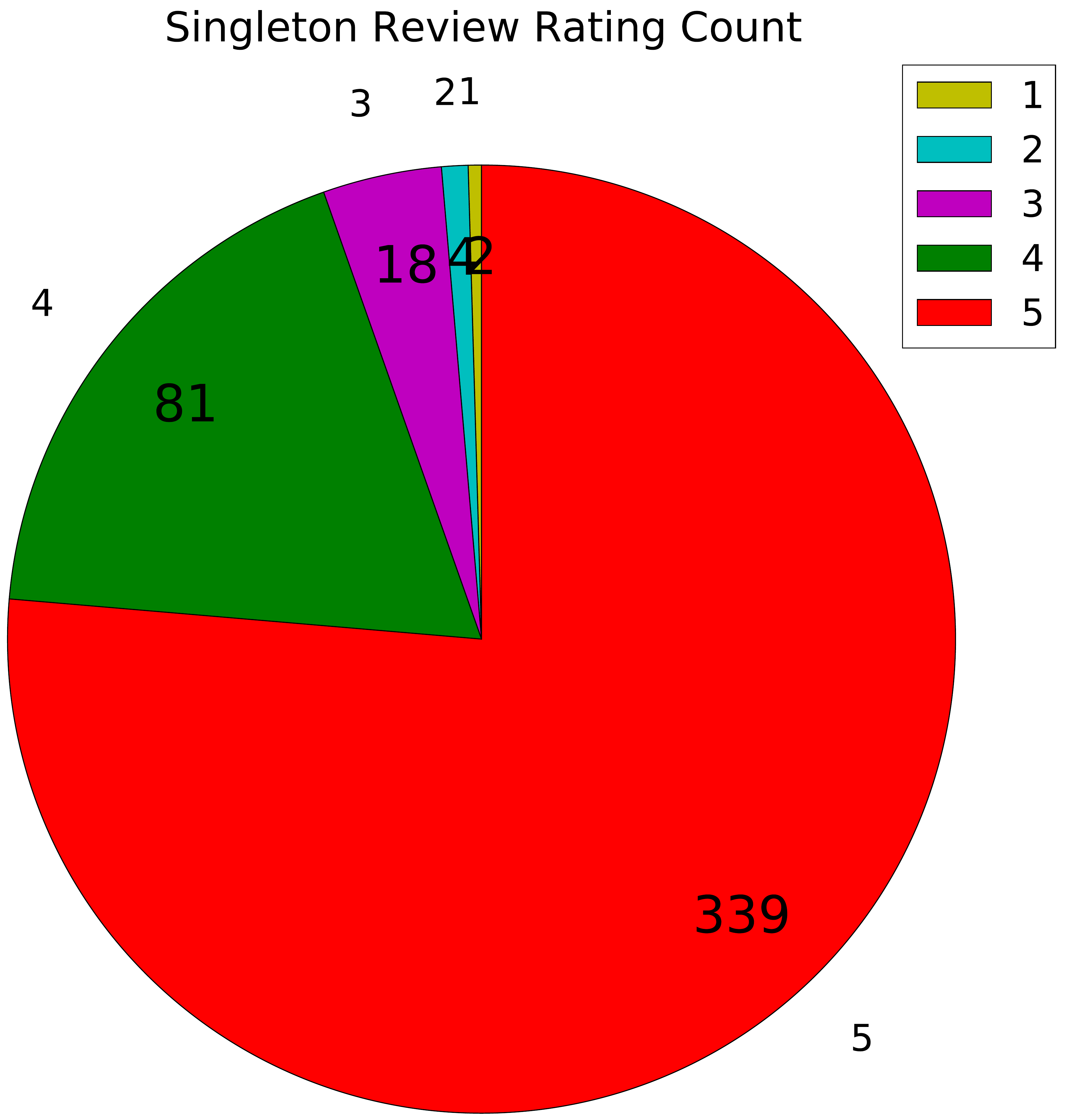} &
		\hspace{0.1in}	\includegraphics[width=1.3in,height=1.3in]{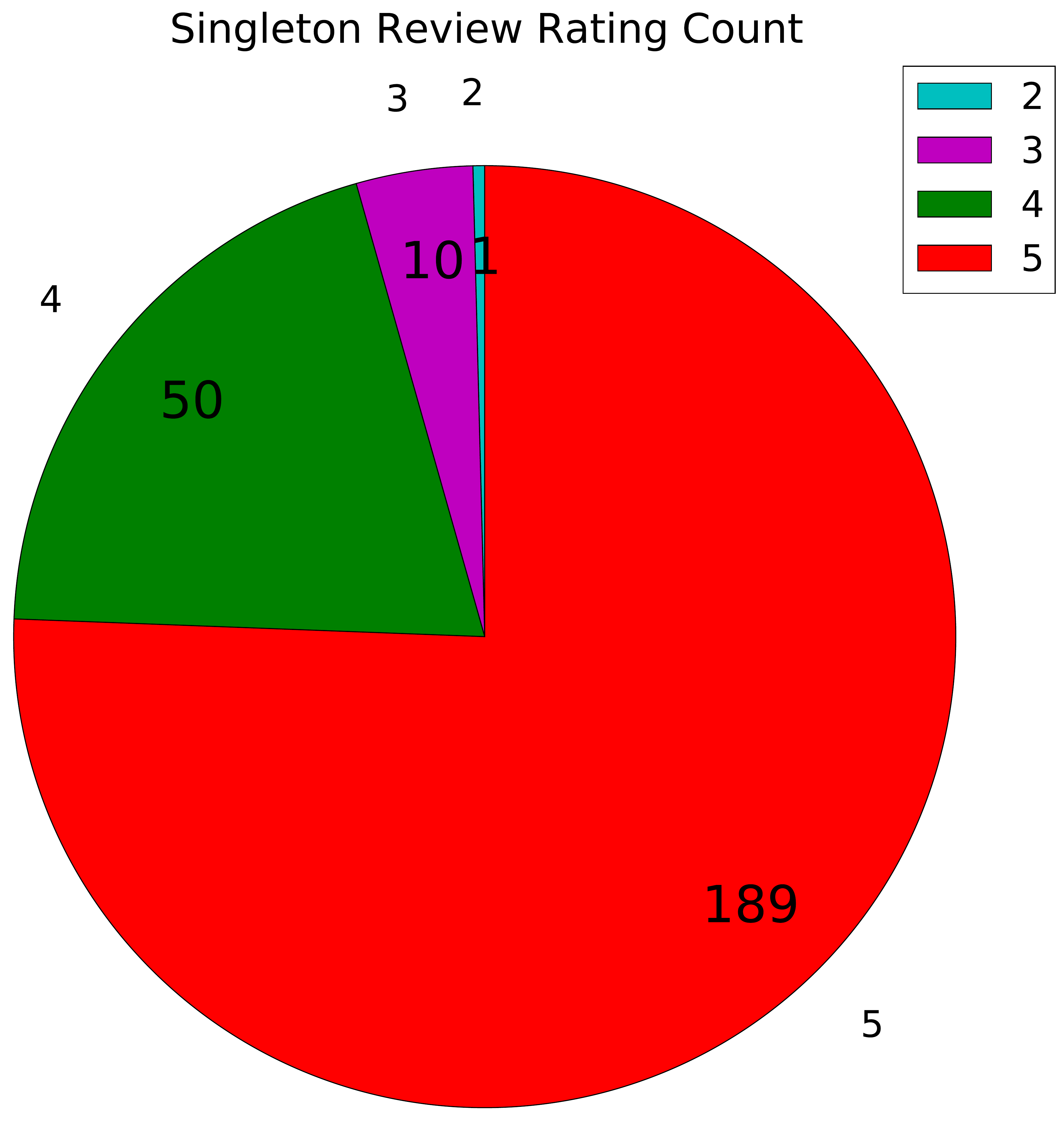} \\
	\end{tabular}
	\vspace{-0.15in}
	\caption{\em \small
		Distribution of ratings by singleton reviewers only, during the 4 weeks of spam campaigns detected in Figure \ref{fig:crown}. This shows that the campaigns were carried by newly created accounts who provided a large number of 4- or 5-star ratings. 
	} 
	\label{fig:swmcase1single}
	\vspace{0.05in}
\end{figure}

Figure \ref{fig:swmcase1} shows the daily review counts for the week before, at, and after the spam campaigns, for each of the 4 detected campaigns, along with the distribution of ratings per day.
Most of these reviews are from \textit{singleton} reviewers that rated this app as 4 or 5 stars, as illustrated in Figure \ref{fig:swmcase1single}.

\begin{figure}[h]
	\vspace{-0.15in}
	\centering
	\begin{tabular}{cc}
		\multicolumn{2}{c}{\includegraphics[width=1.7in,height=1.01in]{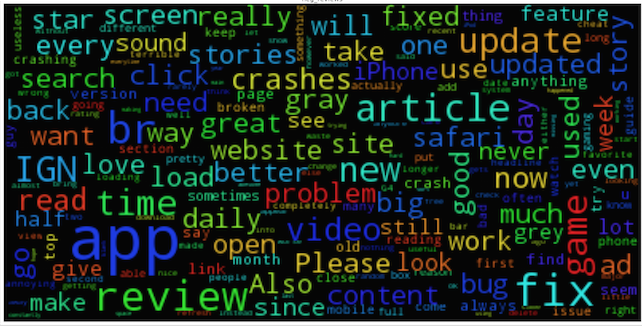}}\\
		\multicolumn{2}{c}{(a) 1-2$\star$ reviews (weeks 75 to 165)}\\
		\hspace{-0.15in}	\includegraphics[width=1.65in,height=1.01in]{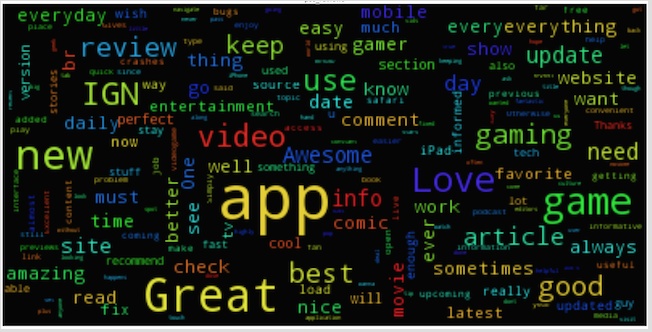} &
		\hspace{-0.1in}	\includegraphics[width=1.65in,height=1.01in]{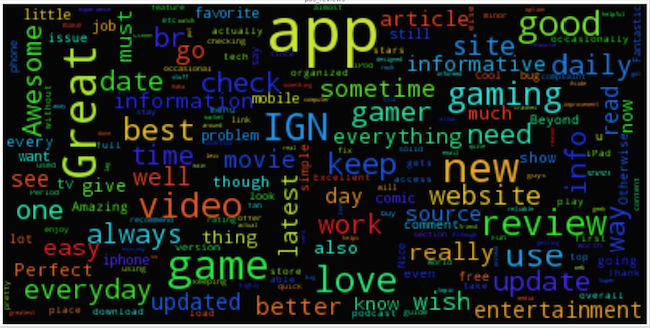} \\
		\hspace{-0.1in}		(b) 4-5$\star$ reviews (week 168) & \hspace{-0.1in} (c) 4-5$\star$ reviews (week 175) \\
		\hspace{-0.15in}	\includegraphics[width=1.65in,height=1.01in]{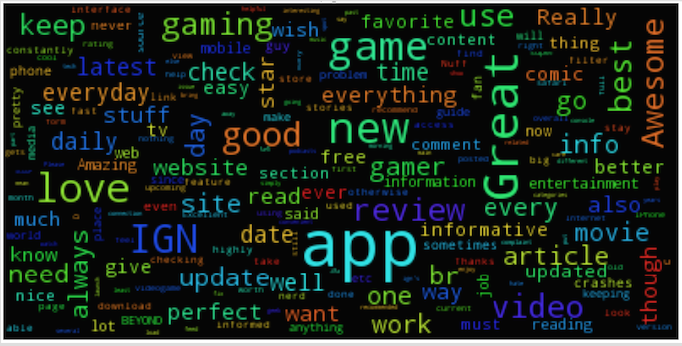} &
		\hspace{-0.1in}	\includegraphics[width=1.65in,height=1.01in]{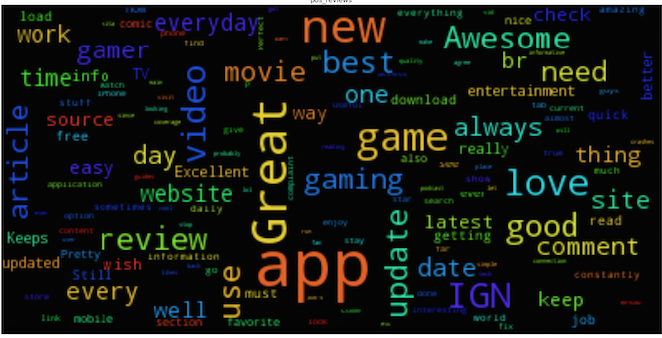} \\
		\hspace{-0.1in}				(d) 4-5$\star$ reviews (week 182) & \hspace{-0.1in} (e) 4-5$\star$ reviews (week 189) \\
	\end{tabular}
	\vspace{-0.15in}
	\caption{\em \small
		(a) Word cloud of negative reviews for time period during which average rating gradually declines for the product 
		in Figure \ref{fig:crown}. (b-e) Word clouds of positive reviews during the respective 4 weeks of spam campaigns.
	} 
	\label{fig:swmcase1text}
	\vspace{0.05in}
\end{figure}

Next we investigate the review text to gain further insight into the campaigns.
First we filter out all the negative (1-2 star) reviews  of this app between weeks 75 to 165, during which the average rating was on a declining trend. Figure \ref{fig:swmcase1text} (a) shows the word-cloud created from these reviews.
It suggests that there is a ``problem'' with the app as it ``crashes'' often and needs a ``fix''.
Next we filter out the positive (4-5 star) reviews from each of the spam campaign weeks, and show the resulting word-clouds in Figure \ref{fig:swmcase1text} (b-e), respectively.
We notice that the new reviewers think the app is ``great'' after the ``new update'' and they ``love'' it.

On further analysis we find that most of the reviews that singletons have written have duplicates or near duplicates.
For example all the following text snippets appear multiple times across different reviews of this app: Great app for gamers,  Great App For Gaming news,  Great app, Easy to use, Must have app for gamers, One of my favorite apps, Use it every day, Very Informative.

\vspace{-0.025in}
\subsubsection{\itunes~Case II: Movie app}
\begin{figure*}[!t]
	%\vspace{-0.15in}
	\centering
	\begin{tabular}{c}
		\includegraphics[width=6.25in,height=4.85in]{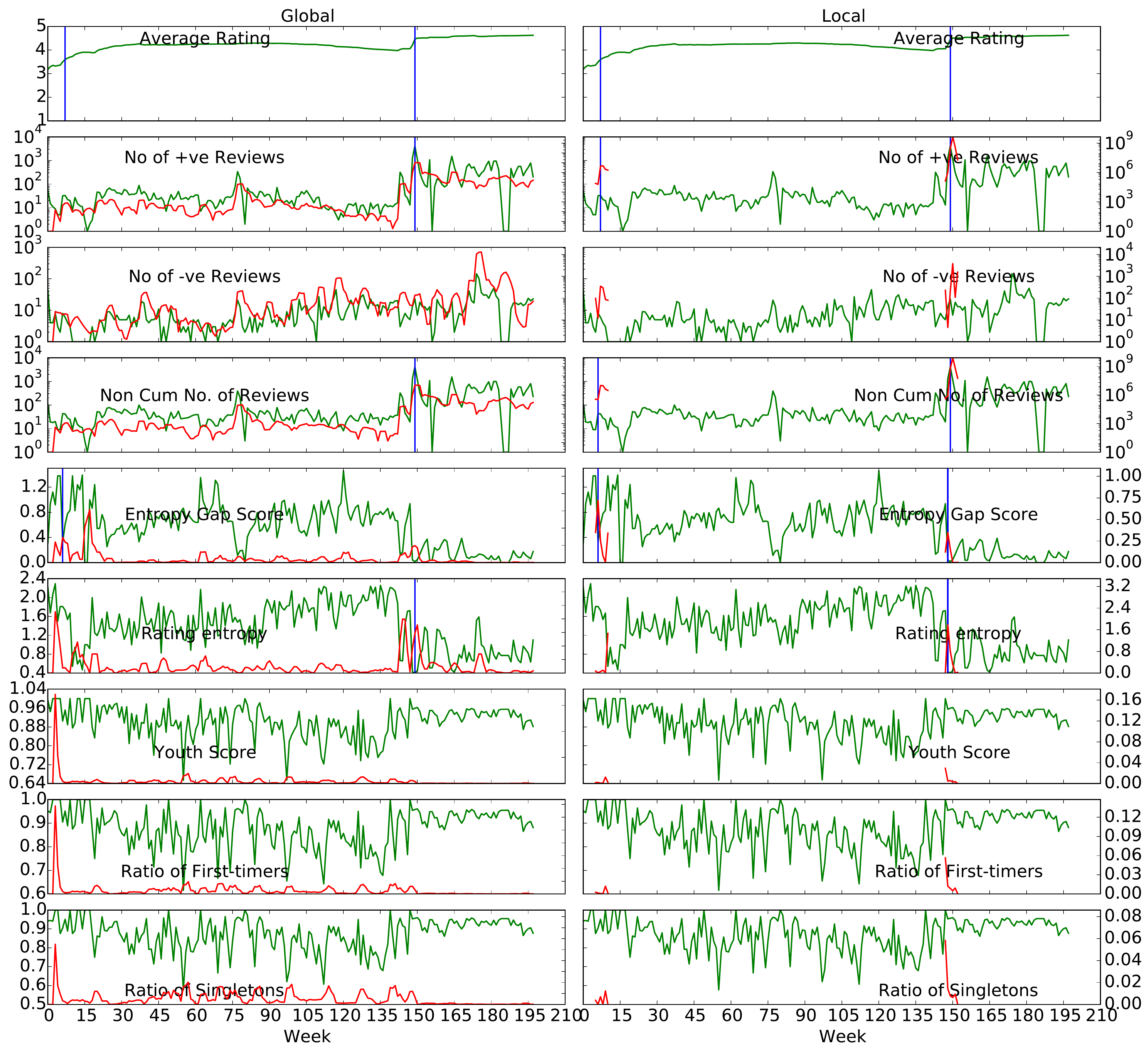} \\
	\end{tabular}
	\vspace{-0.15in}
	\caption{\em \small
		Time series for 9 indicative signals for a software product (app) in \itunes~ (green curves with left y-axis: signal values, red curves with right y-axis: anomaly score, blue bars: anomalous time points). 
		Lead signal: Average Rating (top row). Left: \gar~results, Right: \lar~results. 
		Note that while \gar~computes anomaly scores for all time points for all signals, \lar~computes scores only for those time points indicated by the lead (notice the red curves), being efficient and equally effective.
	} 
	\label{fig:swmcase2time}
	\vspace{-0.1in}
\end{figure*}

Our second case study is for a movie app from \itunes.
Figure \ref{fig:swmcase2time} shows the timeline for the 9 indicative signals, where `Average Rating' is designated as the lead.
Week 149 is detected as anomalous at which the average rating increases from around 3.7 to above 4.

We explicity show the anomaly scores (red curves) computed both by \gar~(left) and \lar~(right) and the detected anomalous points (blue bars) for comparison.
We notice that the anomalies detected by both methods are comparable, while \lar~computes scores only for those points around the anomalies indicated by the lead, being more efficent and as effective. 

\begin{figure}[!t]
	\centering
	\begin{tabular}{cc}
		\multicolumn{2}{c}{	\hspace{-0.1in}\includegraphics[width=3.15in,height=1.2in]{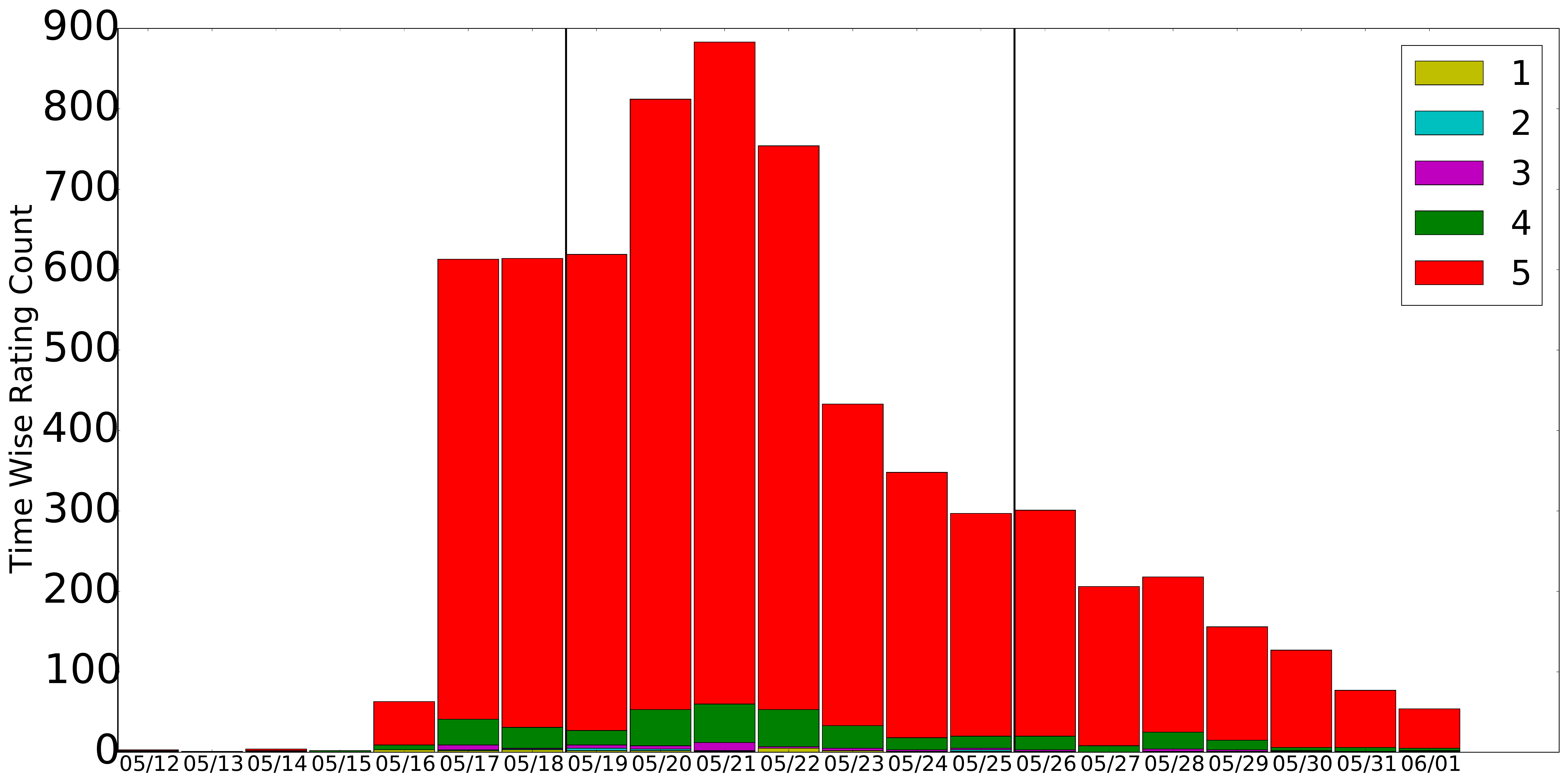}} \\\\
		\hspace{0.0in}\includegraphics[width=1.45in,height=1.45in]{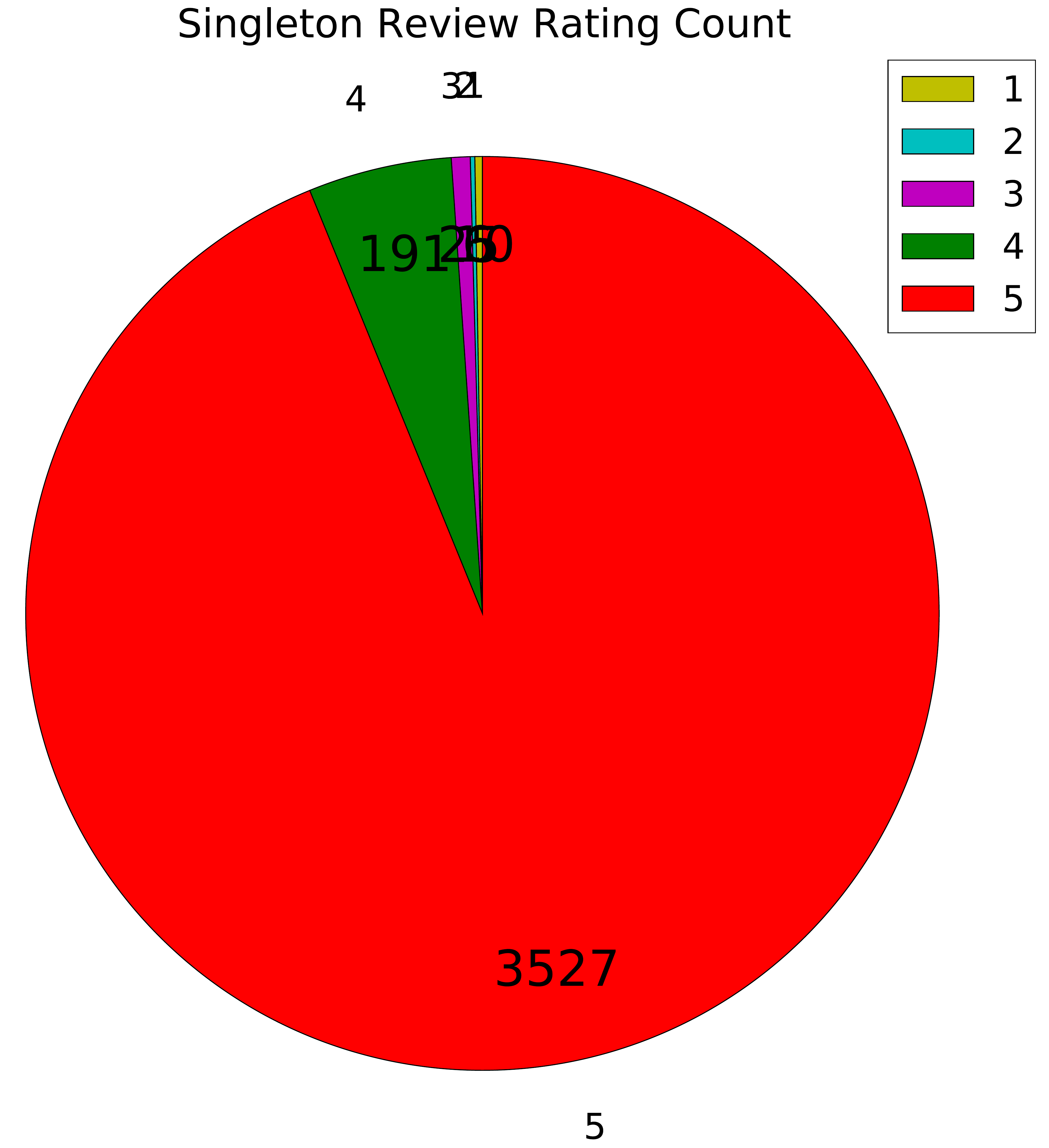} &
		\hspace{0.0in}	\includegraphics[width=1.45in,height=1.45in]{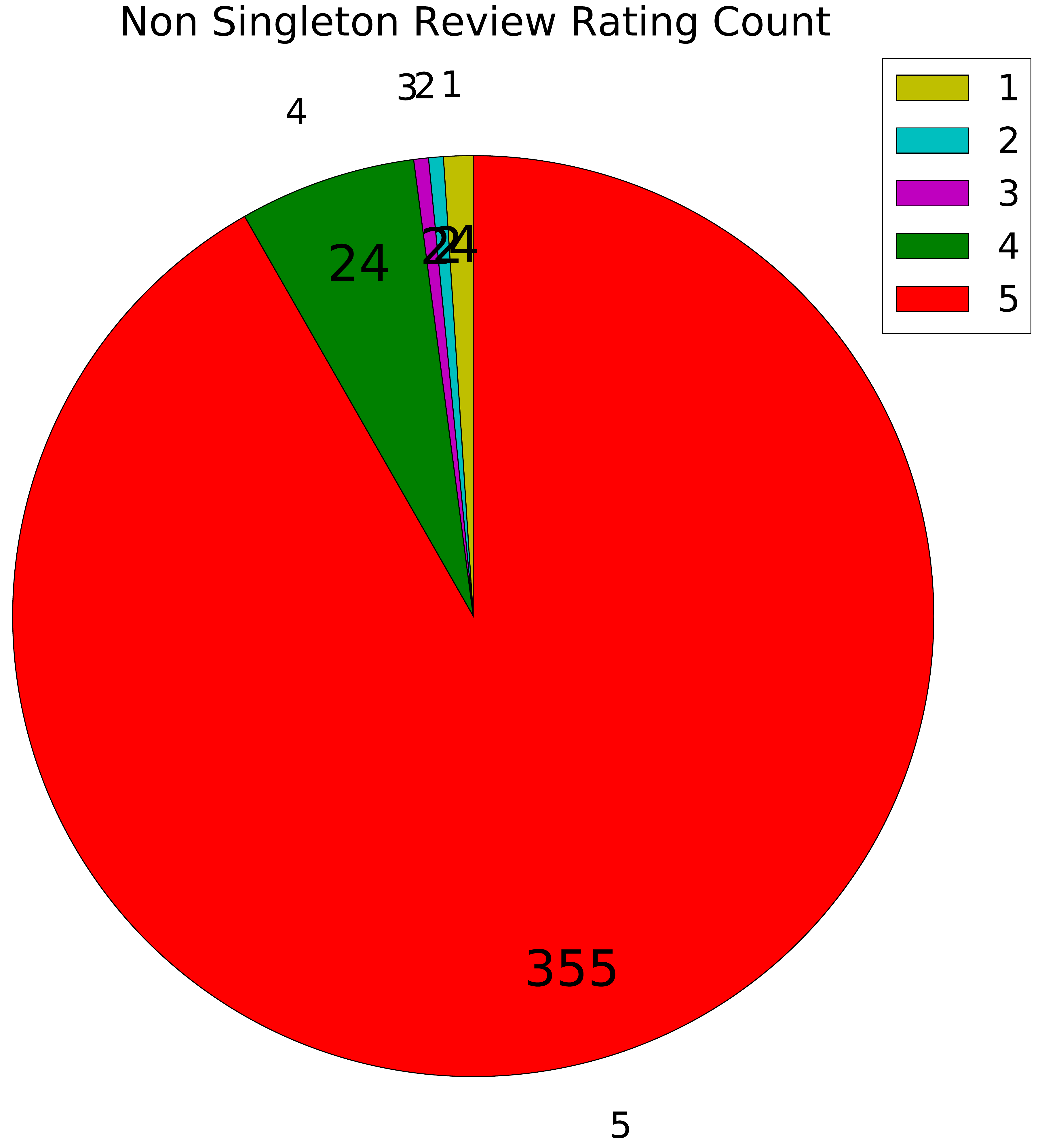} \\
	\end{tabular}
	\vspace{-0.15in}
	\caption{\em \small
		(top) Daily review counts before, during, and after the spam campaign detected on week 149  in Figure \ref{fig:swmcase2time}.
		$>$4,000 reviews during campaign, highest review count per day: $\approx$900.
		(bottom) 
		Rating distribution of (majority of) reviews from singleton (left) and non-singleton reviewers (right) in week 149. 
	} 
	\label{fig:swmcase2line}
\end{figure}

Figure \ref{fig:swmcase2line} (top) shows the daily review counts, stacked by star-rating, before, during, and after the campaign week. This campaign is voluminous: a total of more than 4,000 reviews are written in week 149.
The highest number of reviews per day is around 900 (May 21), which is roughly 1 review for every 2 minutes if the reviews were written by a single user.
The piecharts in Figure \ref{fig:swmcase2line} show the rating distribution of reviews from singletons (left) and non-singletons (right) during that week.
Again, most spam reviews are 5-star rated and written from new accounts.

The following text is shared among singleton reviewers of this product at various weeks near the anomalous week: Best movie app ever!!,  Best movie app out there., Best movie app out there!,  I use it all the time,  I use this all the time!,  I use this app all the time., I use this app all the time!, Love it!, One of my favorite apps!!!, Love the app, use it all the time!,  Way better than Fandango!.

\vspace{-0.015in}
\subsubsection{\itunes~Case III: Whip-sound app}
Outliers in data need not always correspond to anomalies (in our case, spam campaigns) but also novelty or new trends.
We present such an example from the \itunes~dataset.

\begin{figure}[!t]
	\centering
	\begin{tabular}{c}
		\includegraphics[width=3.15in,height=4.75in]{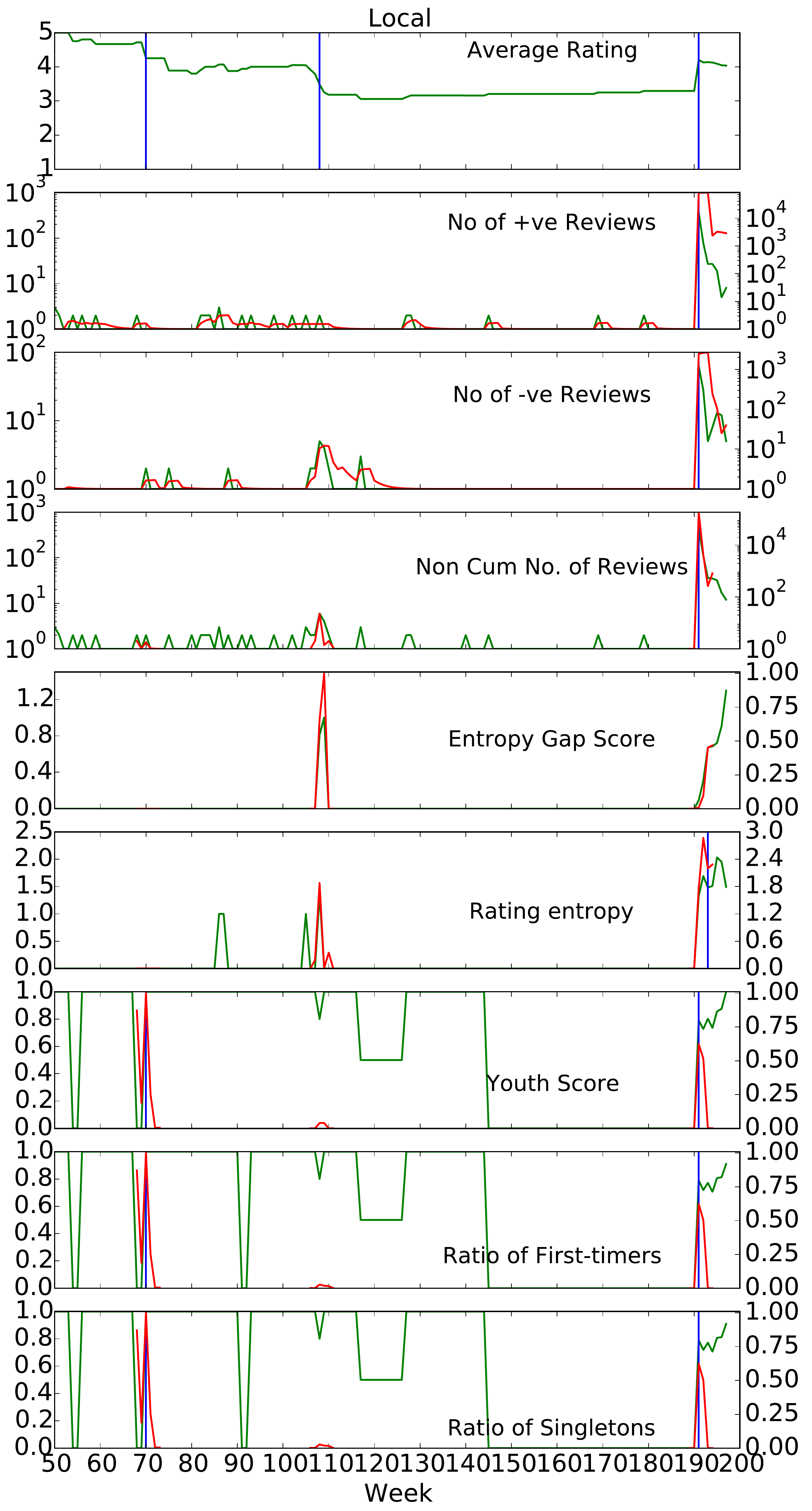} \\
	\end{tabular}
	\vspace{-0.1in}
	\caption{\em \small
		Time series for 9 indicative signals for a whip-sound app in \itunes.
		On week 191, this app appeared in ``Big Bang Theory'' (a popular TV show) Season 5 Episode 19.
	} 
	\label{fig:swmcase3time}
	\vspace{-0.2in}
\end{figure}

\begin{figure}[!h]
	\centering
	\begin{tabular}{c}
		\hspace{-0.2in}\includegraphics[width=2.25in,height=1.25in]{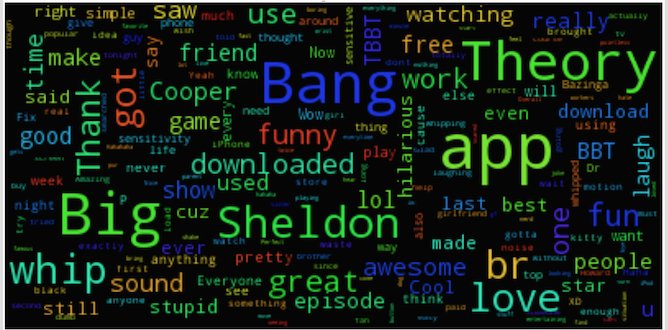} \\
	\end{tabular}
	\vspace{-0.15in}
	\caption{\em \small
		Word cloud of reviews during week 191 in Figure \ref{fig:swmcase3time}.
	} 
	\label{fig:swmcase3word}
	\vspace{-0.1in}
\end{figure}

The product ``Pocket Whip'' is a famous app that produces a whip sound. A TV show called the ``Big Bang Theory'' introduced this whip app in season 5 episode 19. It was aired on March 8, 2012 and consequently, 
reviewers started downloading the app and writing reviews from March 9, 2012 throughout the week.
We detect this activity on week 191 in Figure \ref{fig:swmcase3time} and characterize through Figure \ref{fig:swmcase3word}.
Our approach is powerful enough not only to identify anomalous/suspicious review spam activities but also new trends: e.g., a new version of an app, a new chef in a restaurant, etc.

\begin{figure}[!h]
	\centering
	\begin{tabular}{c}
		\hspace{-0.1in}\includegraphics[width=3.15in,height=1.2in]{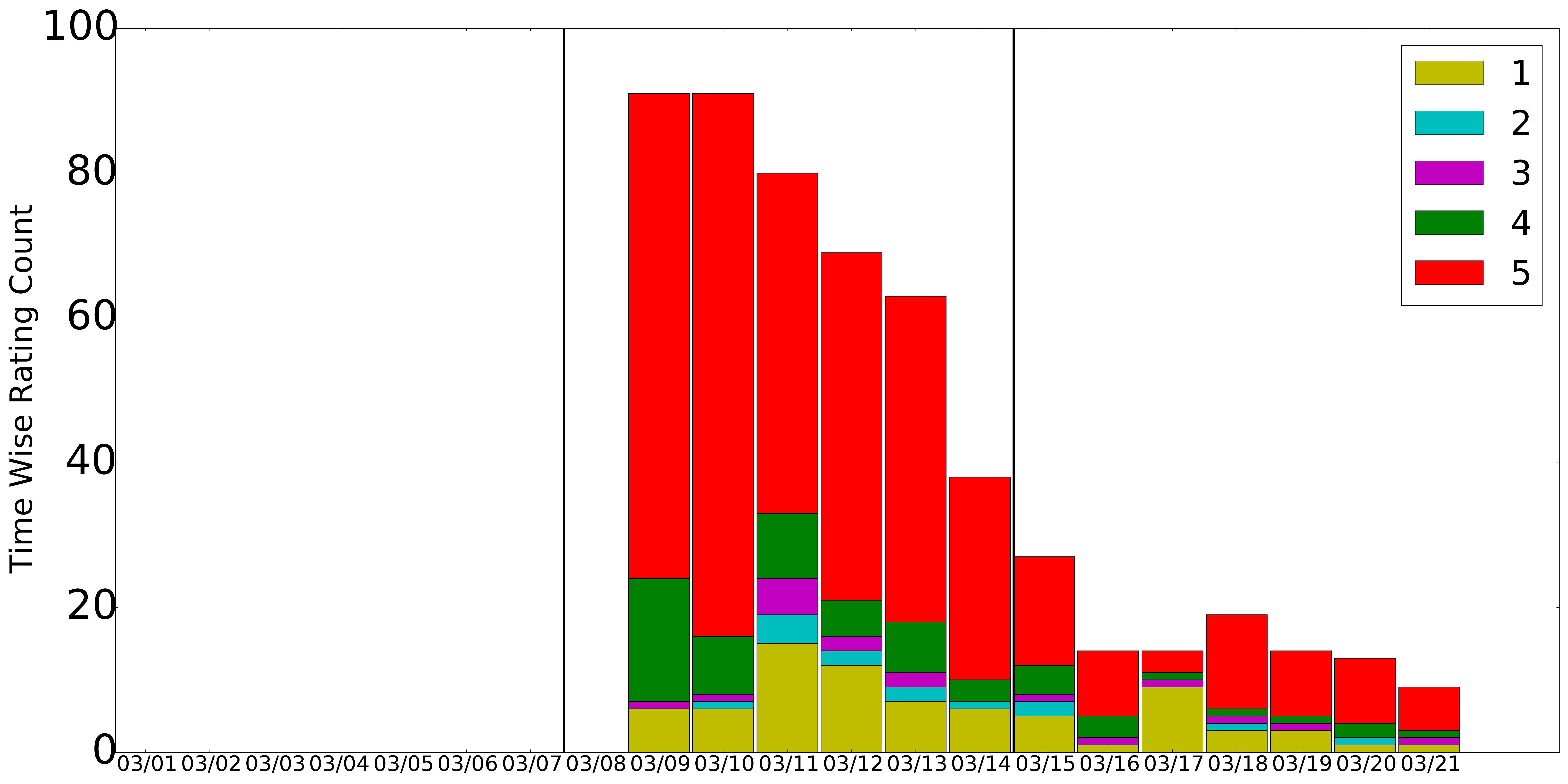} \\
	\end{tabular}
	\vspace{-0.15in}
	\caption{\em \small
		(top) Daily review counts before, during, and after the event detected on week 191  in Figure \ref{fig:swmcase3time}.
	} 
	\label{fig:swmcase3line}
\end{figure}

While the majority of new users liked the app (hence increase in average rating in week 191), number of negative reviews also increased (although not as much, see supporting signal in row 3 of Figure \ref{fig:swmcase3time}). This is also evident from Figure \ref{fig:swmcase3line} that shows the daily rating distributions.

\vspace{-0.015in}
\subsubsection{\flip~Case I}
\begin{figure}[!t]
	%\vspace{-0.15in}
	\centering
	\begin{tabular}{cc}
		\hspace{-0.1in}	\includegraphics[width=1.25in,height=4.35in]{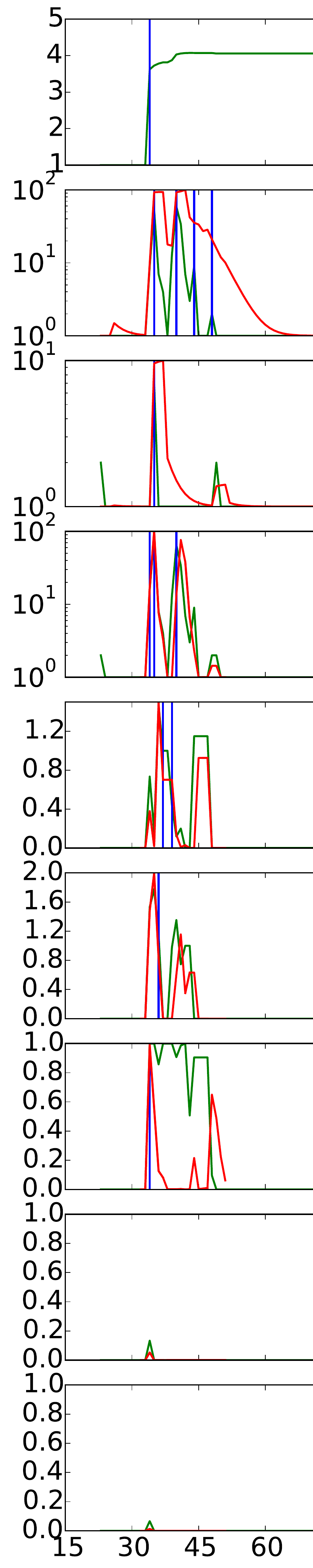} &
		\hspace{0.1in}	\includegraphics[width=1.25in,height=4.35in]{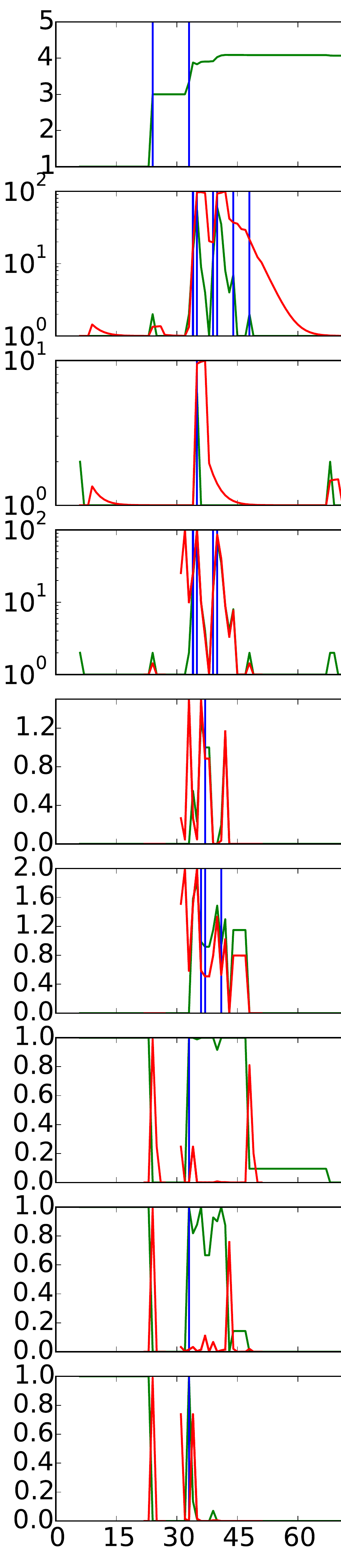}\\
	\end{tabular}
	\vspace{-0.125in}
	\caption{\em \small
		Partial time series for 9 indicative signals for two different products from \flip~that were spammed by same reviewers during same time periods (week 35 and 40).
	} 
	\label{fig:flipcase1}
	\vspace{0.1in}
\end{figure}

\begin{figure}[!h]
	%\vspace{-0.15in}
	\centering
	\begin{tabular}{c}
		\hspace{-0.0in}\includegraphics[width=3.0in,height=1.01in]{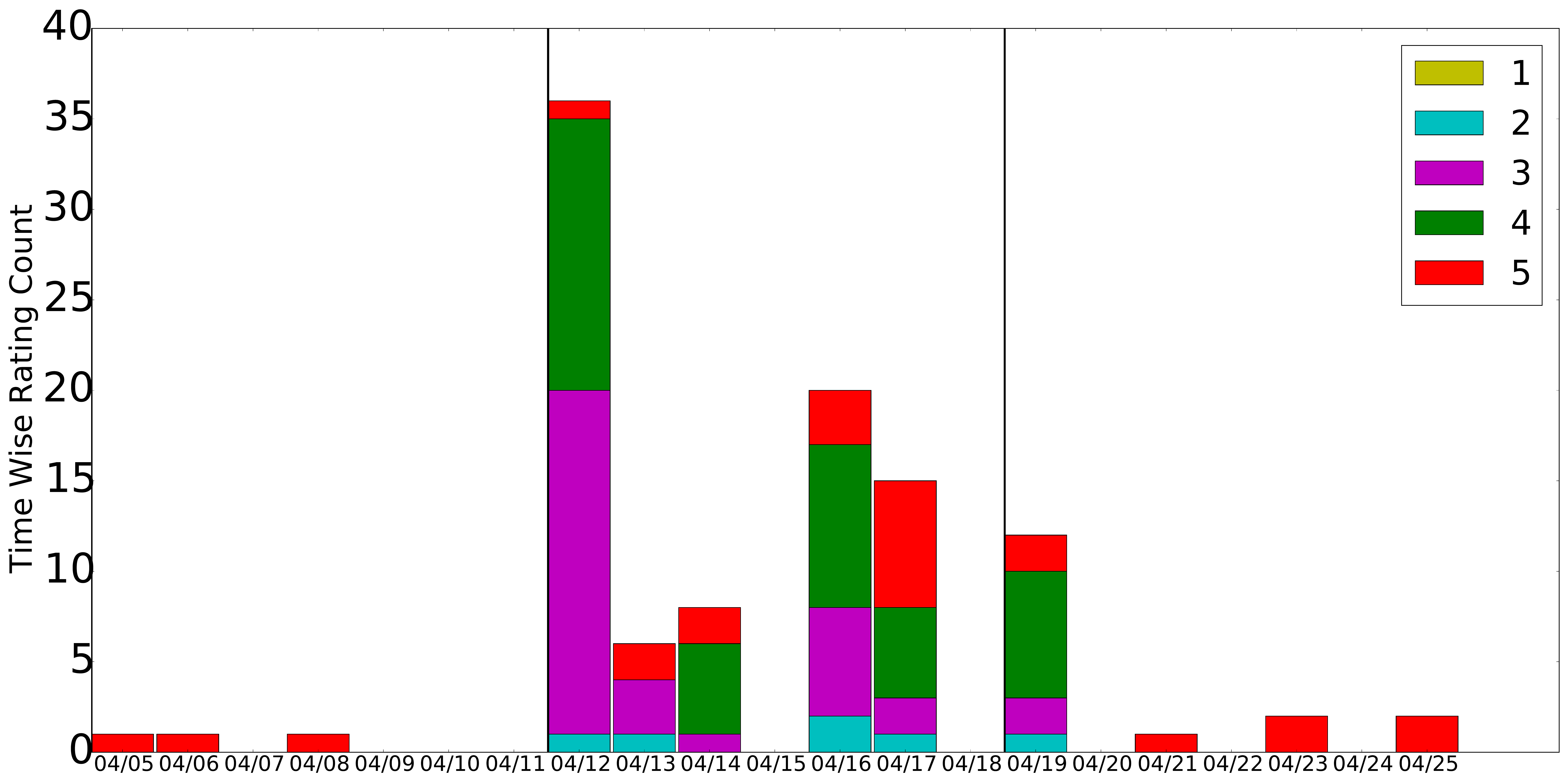} \\
		\hspace{-0.0in}\includegraphics[width=3.0in,height=1.01in]{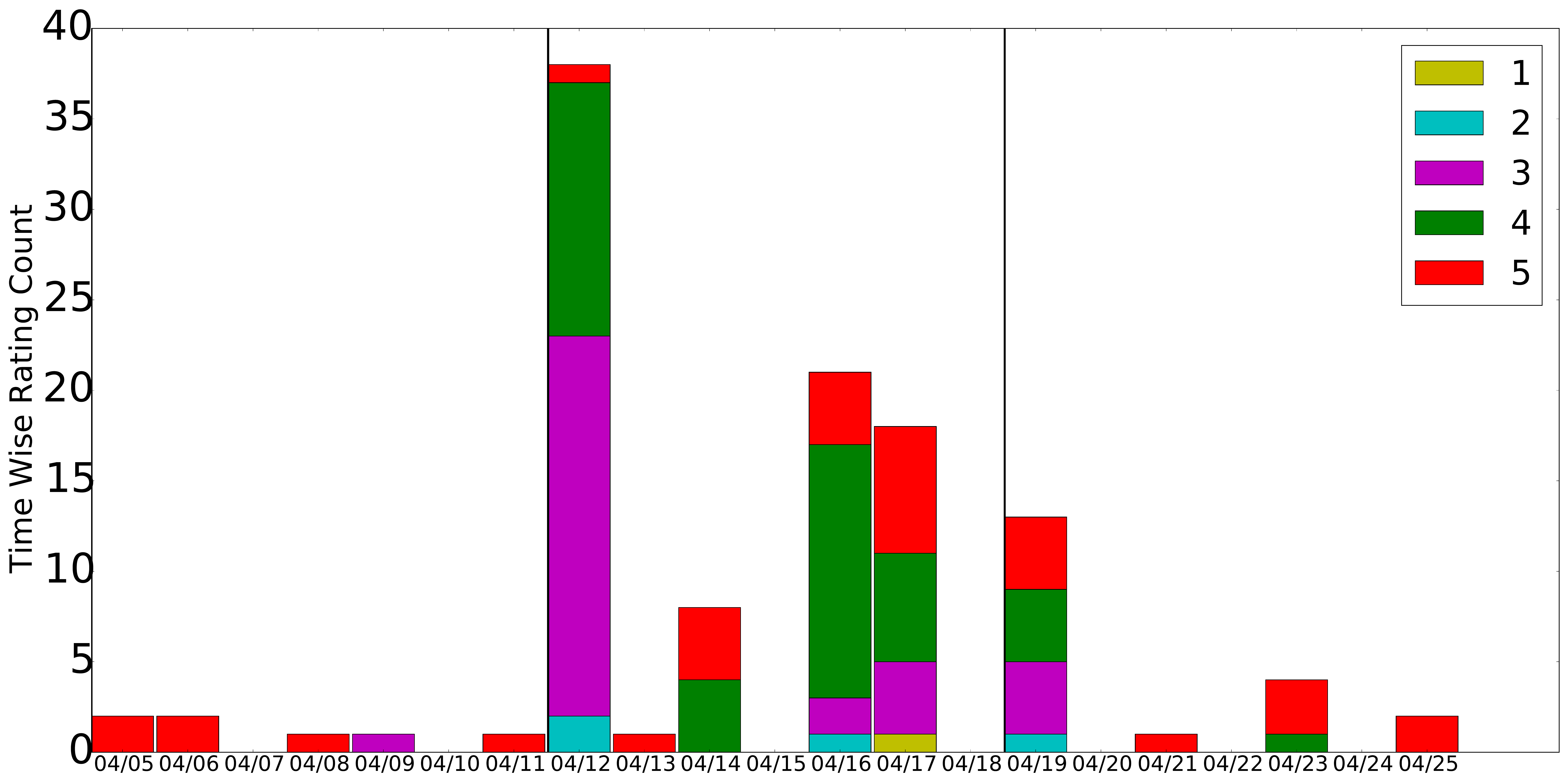} \\
	\end{tabular}
	\vspace{-0.15in}
	\caption{\em \small
		Daily review counts before, during, and after the spam campaign week 35 for two \flip~products in Figure \ref{fig:flipcase1}.
	} 
	\label{fig:flipcase1line}
	\vspace{-0.1in}
\end{figure}

In Figure \ref{fig:flipcase1} (left) we show one of the most suspicious products in \flip, where week 35 is detected as anomalous. During this period, around 80 reviews are written. However the characteristics of this campaign is different in two aspects.
First, most reviews are rated 3- or 4-stars, but only a few 5-stars (See Figure \ref{fig:flipcase1line} (top)), while being able to increase average rating. These mixed ratings appear to be for better camouflage.
Second, most reviewers are non-singletons (unlike in \itunes) (notice no change in Ratio of First-timers and Singletons) although they are young accounts (notice spike in Youth Score).
This suggests that other (related) products might also have been spammed by the same reviewers.

\begin{figure}[!h]
	%\vspace{-0.15in}
	\centering
	\begin{tabular}{c}
		\hspace{-0.0in}\includegraphics[width=2.75in,height=1.75in]{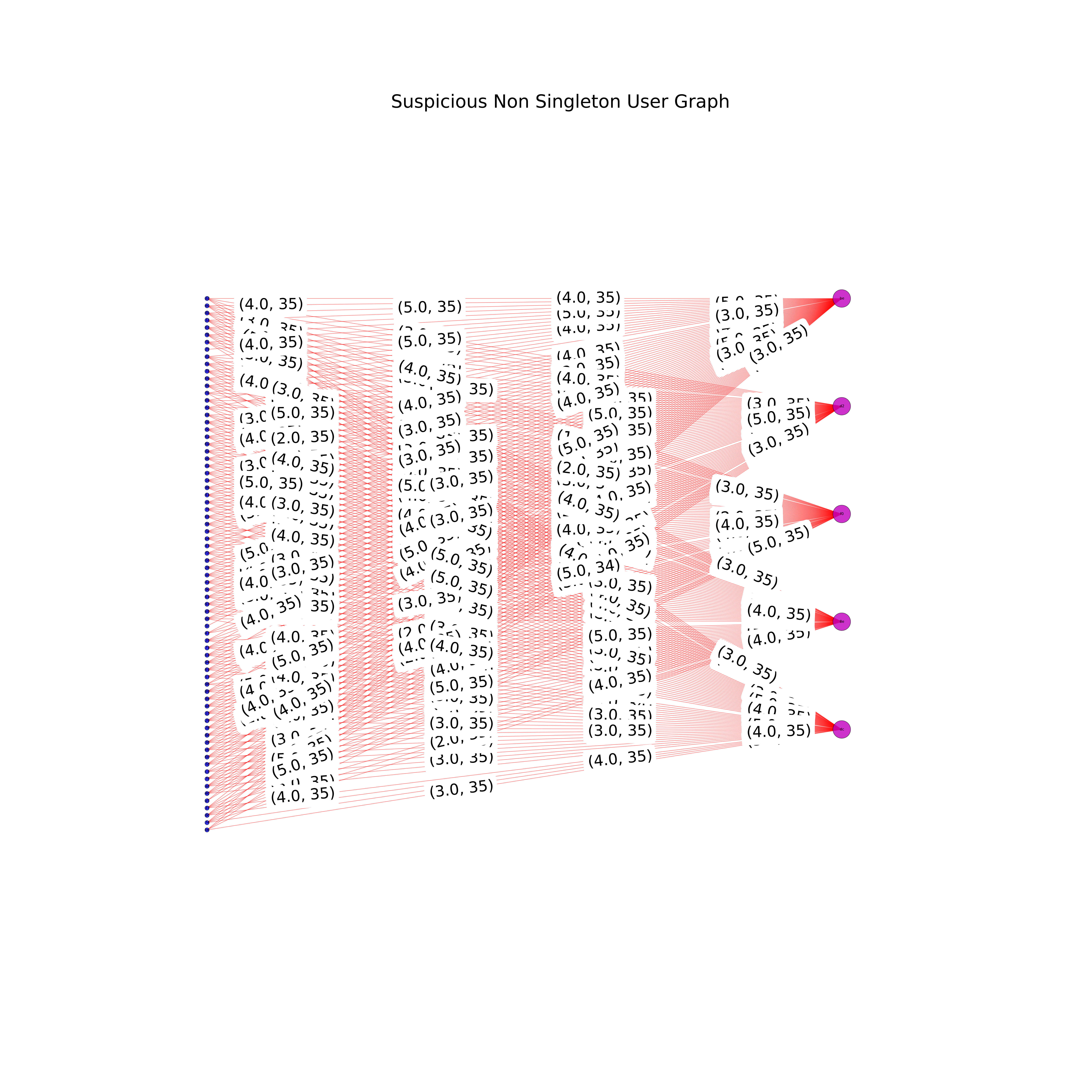} 
	\end{tabular}
	\vspace{-0.15in}
	\caption{\em \small
		Reviewers (left) concurrently spammed multiple hair products (right) from \flip, including those in Figure \ref{fig:flipcase1}. Edge labels: (rating, campaign week)
	} 
	\label{fig:flipcase1thegraph}
	\vspace{-0.1in}
\end{figure}

Figure \ref{fig:flipcase1thegraph} confirms this conjecture, where we find that a list of other products are rated similarly by these users {\em during the same time period}. Moreover, these periods are also detected as anomalous by our method. We show one other example product in Figure \ref{fig:flipcase1} (right) and its coresponding daily review count in Figure \ref{fig:flipcase1line} (bottom)---notice the alignment in time and rating distribution (!)
Further, we find that all these products are hair-related, including straighteners, dryers, and shavers, potentially belonging to the same seller.

\vspace{-0.015in}
\subsubsection{\flip~Case II}
\begin{figure}[!t]
	\centering
	\begin{tabular}{c}
		\includegraphics[width=3.15in,height=2.0in]{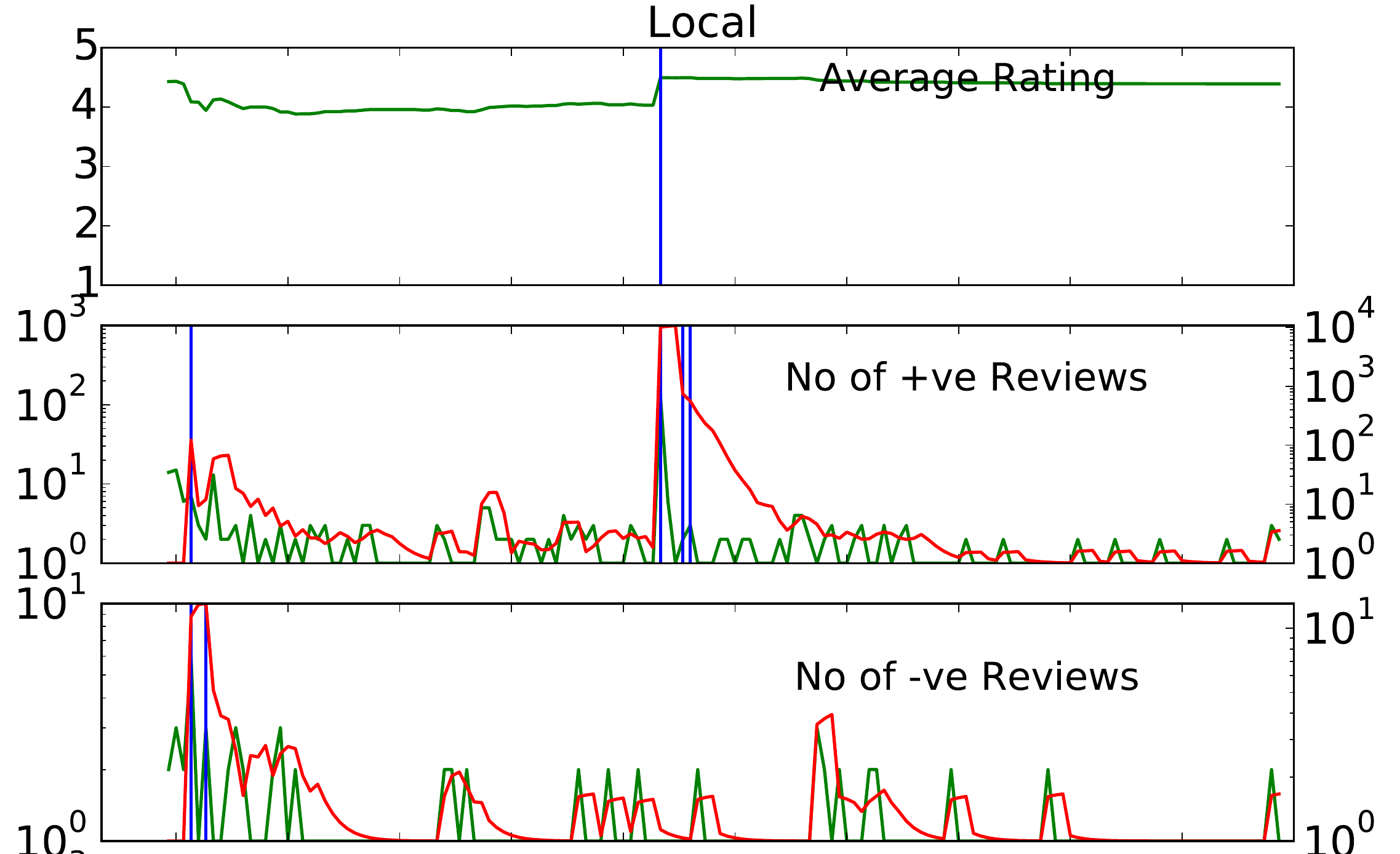} \\
	\end{tabular}
	\vspace{-0.15in}
	\caption{\em \small
		Time series for 3 lead signals for a \flip~book.
	} 
	\label{fig:flipcase2time}
	\vspace{0.1in}
\end{figure}

\begin{figure}[h]
	\centering
	\begin{tabular}{cc}
		\hspace{-0.125in}	\includegraphics[width=2.2in,height=1.1in]{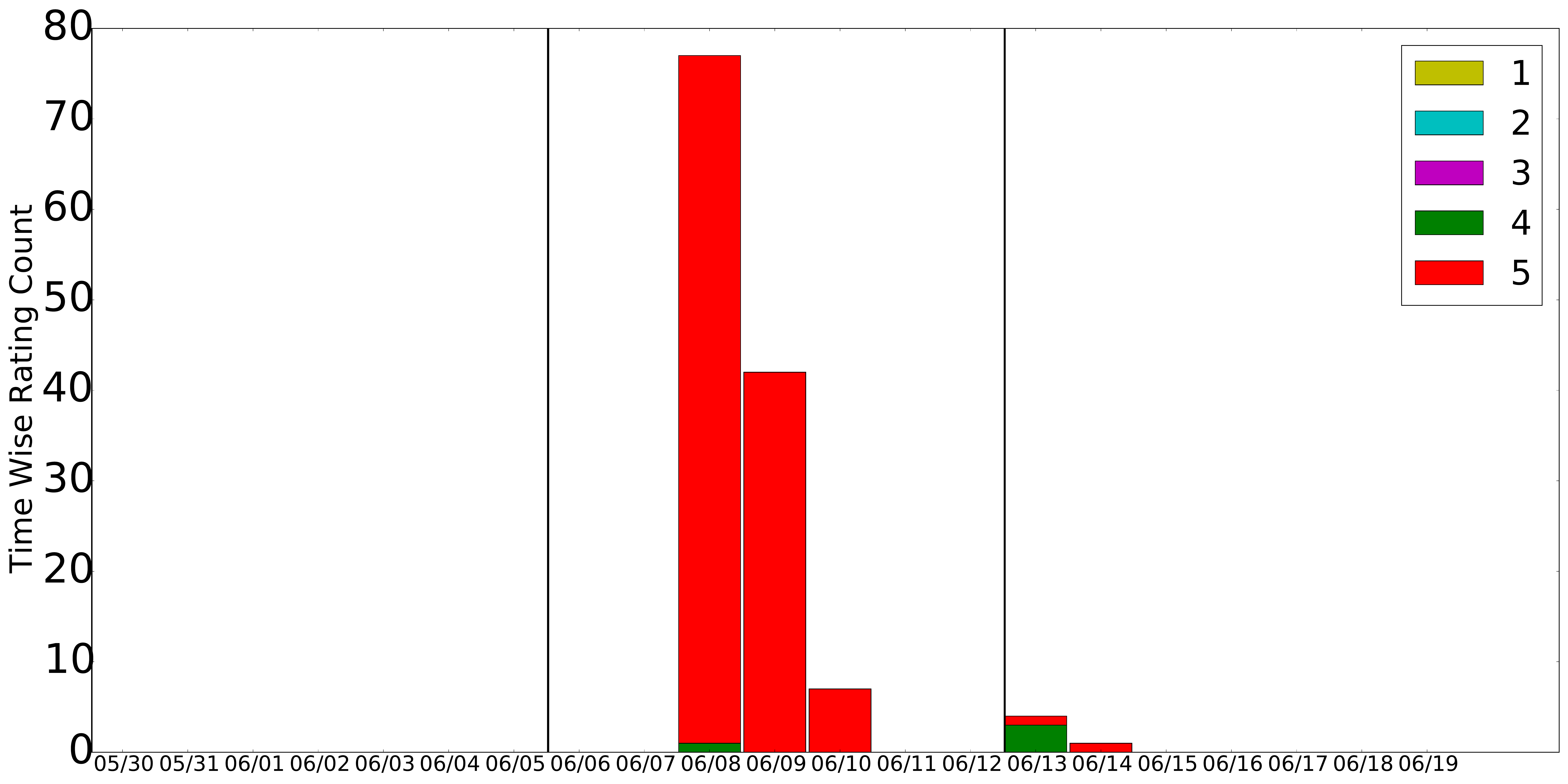} &
		\hspace{-0.1in}	\includegraphics[width=1.1in,height=1.1in]{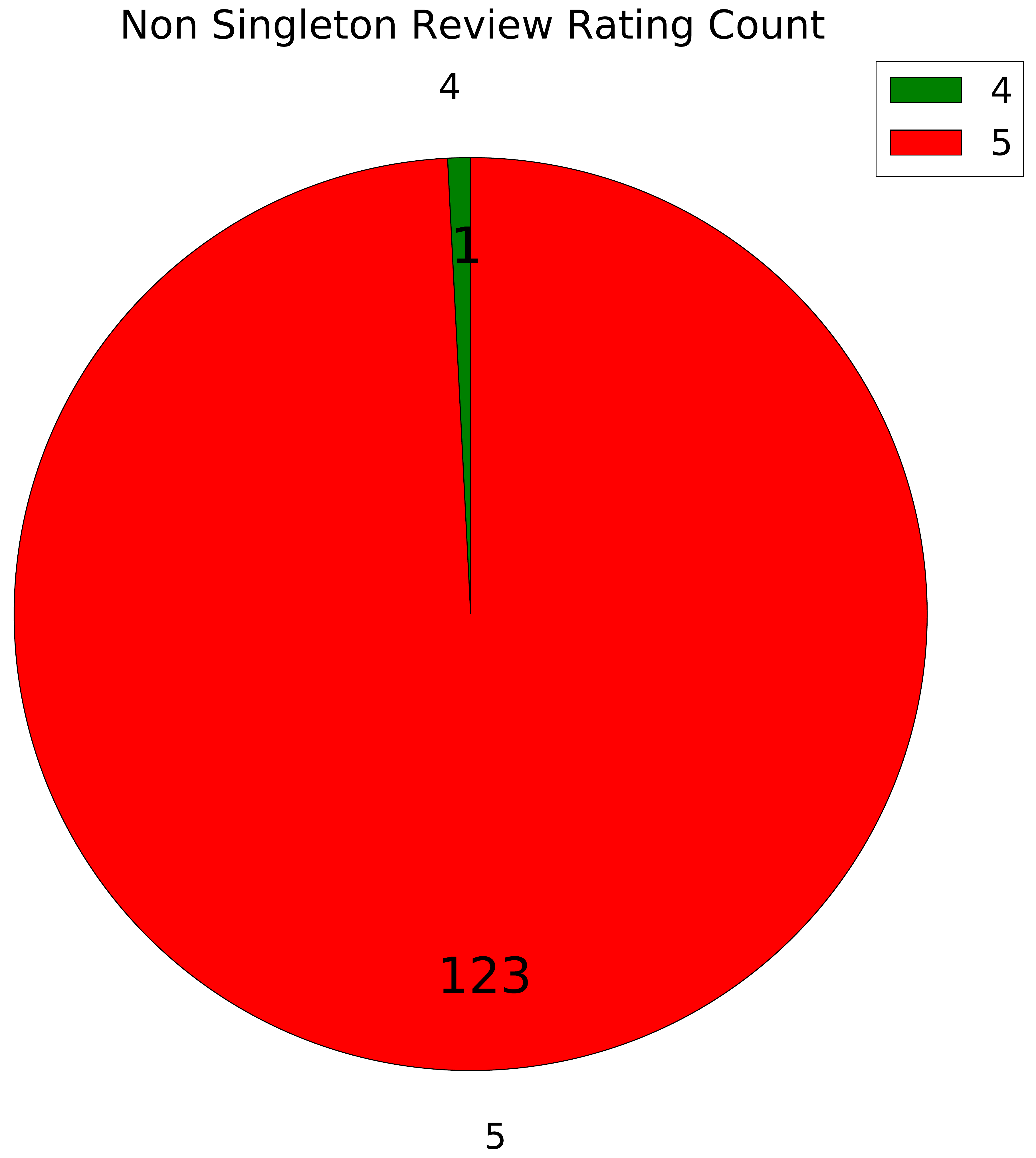} \\
	\end{tabular}
	\vspace{-0.15in}
	\caption{\em \small		(left) Daily review counts, and (right) Rating distribution by non-singletons in week 35 for product in Figure \ref{fig:flipcase2time}.} 
	\label{fig:flipcase2}
	\vspace{0.05in}
\end{figure}

Our second case study is for an anomalous book from \flip.
As shown in Figure \ref{fig:flipcase2time}, its average rating increased to 4.4 on week 95.
We find that this book received 125 5-star reviews in mainly two days during that week.
Surprisingly those were from non-singletons (See Figure \ref{fig:flipcase2}), who also reviewed another product---also a book (!)
Further investigation revealed that
those were 2 out of 3 books of an author.
The one in Figure \ref{fig:flipcase2time} had average rating 3.2
on Goodreads.com.
Similarly, their other book was rated 3.3 on Goodreads while its (potentially spammed)
Flipkart rating is 4.5.
We found that almost all 125 reviewers are \textit{in common}, and write 5-star reviews 7 PM--11.45PM on June 8 and 11AM--7PM on June 9, 2013.  What is more, their reviews follow nearly the \textit{same order} for both books.

\vspace{-0.05in}
\section{Conclusion}
\label{sec:conclude}

Opinion spam has become a prevalent problem, for which a vast body of methods operate in an offline fashion on a collection of static data.
In this work, we brought emphasis to the aspect of time, and approached this problem with a novel temporal formulation.
We proposed a new methodology that (1) monitors a comprehensive list of indicative signals over time, (2) spots anomalous events in real-time, and (3) provides descriptive pointers for manual inspection and characterization. 
As such, our approach exhibits desirable properties, as it is online, efficient, descriptive, and general.

Importantly, while we applied our methodology on opinion spam, it is general enough to be employed for other applications in which multiple signals are monitored over time, such as enterprise security, cyber-physical sensor systems, environmental monitoring, and surveillance systems.

{\footnotesize{
\bibliographystyle{aaai}
\bibliography{BIB/refs}
}}

\end{document}